\documentclass{jfm}

\usepackage{graphicx}
\usepackage{newtxtext}
\usepackage{newtxmath}
\usepackage{natbib}
\usepackage{hyperref}
\hypersetup{
    colorlinks = true,
    urlcolor   = blue,
    citecolor  = black,
}

\newcommand{\RomanNumeralCaps}[1]
\linenumbers

\usepackage{setspace}
\usepackage{afterpage}
\usepackage[per-mode=symbol]{siunitx}
\usepackage{tabulary,graphicx,amsfonts,amsmath,amssymb,amsbsy,dcolumn,bm}
\usepackage[utf8]{inputenc}
\usepackage[T1]{fontenc}
\usepackage{gensymb}
\usepackage{graphicx}
\usepackage{dcolumn}
\usepackage{bm}
\usepackage{microtype}

\usepackage{xcolor}
\usepackage{hyperref}
\usepackage{tabularx} 
\usepackage{booktabs}
\usepackage{widetable}
\usepackage{float}
\usepackage{siunitx}
\usepackage{amsmath}

\newlength{\toprulewidth}
\patchcmd{\toprule}
  {\heavyrulewidth}{\toprulewidth}
  {}{}

\hypersetup{
  colorlinks,
  citecolor=green,
  linkcolor=red,
  urlcolor=blue}

\usepackage{url,multirow,morefloats,floatflt,cancel,tfrupee}
\makeatletter

\AtBeginDocument{\@ifpackageloaded{textcomp}{}{\usepackage{textcomp}}}
\makeatother
\usepackage{colortbl}
\usepackage{xcolor}
\usepackage{pifont}
\usepackage[nointegrals]{wasysym}
\makeatletter

\def\mcWidth#1{\csname TY@F#1\endcsname+\tabcolsep}

\def\cAlignHack{\rightskip\@flushglue\leftskip\@flushglue\parindent\z@\parfillskip\z@skip}
\def\rAlignHack{\rightskip\z@skip\leftskip\@flushglue \parindent\z@\parfillskip\z@skip}

\@ifundefined{etal}{}{}

\usepackage{ifxetex}
\ifxetex\else\if@twocolumn\@ifpackageloaded{stfloats}{}{\usepackage{dblfloatfix}}\fi\fi

\AtBeginDocument{
\expandafter\ifx\csname eqalign\endcsname\relax
\def\eqalign#1{\null\vcenter{\def\\{\cr}\openup\jot\m@th
  \ialign{\strut$\displaystyle{##}$\hfil&$\displaystyle{{}##}$\hfil
      \crcr#1\crcr}}\,}
\fi
}

\AtBeginDocument{%
  \@ifpackageloaded{endfloat}%
   {\renewcommand\efloat@iwrite[1]{\immediate\expandafter\protected@write\csname efloat@post#1\endcsname{}}}{\newif\ifefloat@tables}%
}%

\def\BreakURLText#1{\@tfor\brk@tempa:=#1\do{\brk@tempa\hskip0pt}}
\let\lt=<
\let\gt=>
\def\processVert{\ifmmode|\else\textbar\fi}

\@ifundefined{subparagraph}{
\def\subparagraph{\@startsection{paragraph}{5}{2\parindent}{0ex plus 0.1ex minus 0.1ex}%
{0ex}{\normalfont\small\itshape}}%
}{}

\newcommand\role[1]{\unskip}
\newcommand\aucollab[1]{\unskip}
\usepackage[]{changes}

\@ifundefined{tsGraphicsScaleX}{\gdef\tsGraphicsScaleX{1}}{}
\@ifundefined{tsGraphicsScaleY}{\gdef\tsGraphicsScaleY{.9}}{}
\def\checkGraphicsWidth{\ifdim\Gin@nat@width>\linewidth
	\tsGraphicsScaleX\linewidth\else\Gin@nat@width\fi}

\def\checkGraphicsHeight{\ifdim\Gin@nat@height>.9\textheight
	\tsGraphicsScaleY\textheight\else\Gin@nat@height\fi}

\def\fixFloatSize#1{}
\let\ts@includegraphics\includegraphics

\def\inlinegraphic[#1]#2{{\edef\@tempa{#1}\edef\baseline@shift{\ifx\@tempa\@empty0\else#1\fi}\edef\tempZ{\the\numexpr(\numexpr(\baseline@shift*\f@size/100))}\protect\raisebox{\tempZ pt}{\ts@includegraphics{#2}}}}

\AtBeginDocument{\def\includegraphics{\@ifnextchar[{\ts@includegraphics}{\ts@includegraphics[width=\checkGraphicsWidth,height=\checkGraphicsHeight,keepaspectratio]}}}

\DeclareMathAlphabet{\mathpzc}{OT1}{pzc}{m}{it}

\def\URL#1#2{\@ifundefined{href}{#2}{\href{#1}{#2}}}

\def\UrlOrds{\do\*\do\-\do\~\do\'\do\"\do\-}%
\g@addto@macro{\UrlBreaks}{\UrlOrds}

\edef\fntEncoding{\f@encoding}

\makeatother

\newif\ifmultipleabstract\multipleabstractfalse%
\newif\iffinalmode
\finalmodetrue  

\newcommand{\myadded}[1]{%
  \iffinalmode
    #1%
  \else
    \added{#1}%
  \fi
}

\newcommand{\mydeleted}[1]{%
  \iffinalmode
  \else
    \deleted{#1}%
  \fi
}

\newcommand{\myreplaced}[2]{%
  \iffinalmode
    #1%
  \else
    \replaced{#1}{#2}%
  \fi
}

\newcommand{\myaddedred}[1]{%
  \iffinalmode
    #1%
  \else
    {\color{red}#1}%
  \fi
}

\newcommand{\mydeletedred}[1]{%
  \iffinalmode
  \else
    {\color{red}\sout{#1}}%
  \fi
}

\newcommand{\mydeletedappendix}[1]{%
  \iffinalmode
  \else
    \section{\mydeletedred{#1}}%
  \fi
}

\title{Rim destabilization and re-formation upon severance from its expanding sheet}

\author{M. Kharbedia\aff{1}
B. Liu\aff{1,2}
\mbox{R.A. Meijer}\aff{1,2}
\mbox{D.J. Engels}\aff{1,2}
\mbox{H.K. Schubert}\aff{1,2}
\mbox{L. Bourouiba}\aff{3}
\corresp{\email{lbouro@mit.edu}}
\mbox{O.O. Versolato}\aff{1,2}
\corresp{\email{versolato@arcnl.nl}}}

\affiliation{\aff{1}Advanced Research Center for Nanolithography (ARCNL), Science Park 106, 1098 XG Amsterdam, The Netherlands
\aff{2}LaserLab, Department of Physics and Astronomy, Vrije Universiteit Amsterdam, De Boelelaan 1100, 1081 HV Amsterdam, The Netherlands
\aff{3}The Fluid Dynamics of Disease Transmission Laboratory, Fluids and Health Network, Massachusetts Institute of Technology, 77 Massachusetts Ave, Cambridge, MA 02139, United States
}

\begin{document}
\maketitle
\begin{abstract}

Upon radial liquid sheet expansion, a bounding rim forms, with a thickness and stability governed, in part, by the liquid influx from the unsteady connected sheet. 
We examine how the thickness and fragmentation of such a radially expanding rim change upon its severance from its sheet, absent of liquid influx. 
To do so, we design an experiment enabling the study of rims pre and post severance by vaporizing the thin neck connecting the rim. 
\myadded{No vaporization occurs of the bulk rim itself. }
We confirm that the severed rim follows a ballistic motion, with a radial velocity inherited from the sheet at severance time. 
We identify that the severed rim undergoes fragmentation in two types of junctions: the base of inherited, pre-severance, ligaments and the junction between nascent rim corrugations, with no significant distinction between the two associated timescales. 
The number of ligaments and fragments formed is captured well by the theoretical prediction of rim corrugation and ligament wavenumbers established for unsteady expanding sheets upon droplet impact on surfaces of comparable size to the droplet\mydeleted{, and with the sheet thickness profiles in both systems having the same functional form}. 
Our findings are robust to changes in impacting laser energy and initial droplet size. 
Finally, we report and analyze the re-formation of the rim on the expanding sheet and propose a prediction for its characteristic corrugation timescale. 
Our findings highlight the fundamental mechanisms governing interfacial destabilization of connected fluid-fed expanding rims that become severed, thereby clarifying destabilization of freely radially expanding toroidal fluid structures absent of fluid influx.

\end{abstract}\def\keywordstitle{Keywords}

\maketitle

\section{Introduction}\label{Introduction}

Liquid droplets are ubiquitous and their formation is inevitable in numerous physical processes. 
In any hydrodynamic system where fluid fragmentation occurs, droplets constitute an important liquid reservoir.
They are responsible for microscale mass transport, such as the formation of bio- and atmospheric aerosols\,\protect\citep{lhuissier_bursting_2012,Veron2015,poulain2018ageing,poulain2018biosurfactants,BARBME2021}, turbulent distribution of pollutants in oceans and rivers\,\citep{VERED202160}, and the spread of spores and debris in nature driven by wind and rain splash\,\protect\citep{GB2014,Gilet2015,Lejeuneetal2018}. 
In addition, droplets are a subject of particular interest in industries involving fragmentation and heat transfer, e.g., fuel combustion\,\protect\citep{betelin2012evaporation}, spray coating technology for the deposition of complex liquids on surfaces\,\protect\citep{kumar2023deposition}, ink-jet printing\,\protect\citep{Lohse_inkjet_2022,Liu2023_Inkjet}, or microfluidic devices\,\citep{nielsen20203d, zhang2020acoustic}. 
Distinct physical mechanisms can lead to droplet fragmentation, from direct impact on solid surfaces\myreplaced{\,\mbox{\protect\citep{wang2021droplet, broomWater2022}}}{\mbox{\protect\citep{wang_bourouiba_energypartition, wang_bourouiba_2018_fragspeed, wang_bourouiba_2021_growth,BARFM2021, wang2021droplet, broomWater2022}}}, to atomization of cylindrical jets flowing through micronozzles\,\protect\citep{Holtze_2013}, or laser-driven bursting\,\protect\citep{Gelderblom2016,reijers2017, klein_drop_2020}. 
Particular attention is devoted to microdroplet dynamics in industrial sources of extreme ultraviolet (EUV) light, where nanosecond-pulsed lasers are used to generate EUV light, centered at 13.5\,nm wavelength, from tin microdroplets in two steps: a first \emph{prepulse} laser shapes the droplet into a thin sheet and a second \emph{main pulse} ablates the tin producing EUV light from the resulting plasma\,\citep{jansson_liquid-tin-jet_2004, schupp2019_efficient-EUV, versolato2019physics, Behnke:21}. 
The transformation of the droplet into a sheet leads to a complex mass distribution, evolving into a thinning sheet with a bounding thick rim which subsequently destabilizes into protruded ligaments from which debris is shed\myreplaced{\,\mbox{\citep{villermaux_fragmentation_2007,villermaux_drop_2011, wang_bourouiba_2018_fragspeed, liu_speed_2022}}}{\mbox{\citep{wang_bourouiba_2018_rim, wang_bourouiba_2018_fragspeed, liu_speed_2022, liu_2023_mass}}}. 
The process is unsteady (i.e., the underlying conditions are time-varying), and the rim is a critical link between the sheet and ligaments, while also playing an important role in the temporal evolution of the sheet. 

Prior works on unsteady sheet evolution remark on the key role of the rim \myreplaced{\mbox{\,\citep{villermaux_fragmentation_2007, villermaux_drop_2011, wang_bourouiba_2018_rim}}}{\mbox{\citep{villermaux_fragmentation_2007, villermaux_drop_2011, wang_bourouiba_2018_rim, wang_bourouiba_2021_growth}}} in the context of two canonical examples: The mentioned nanosecond laser and microsized tin droplet interaction and millimeter water droplets impacting solid surfaces\,\citep{villermaux_life_2002, villermaux_fragmentation_2007, villermaux_single-drop_2009, villermaux_drop_2011, wang_bourouiba_2018_rim}. 
Despite the large differences in time, length scales, and impact mechanism, interesting similarities emerged between the two systems with respect to the fundamental mechanisms of rim formation\,\citep{wang_bourouiba_2018_rim}. 
In both cases, it has been proposed that the unsteady sheet and rim dynamics are the consequence of a combination of several factors: the momentum transferred from the impulse (laser or impact) to the droplet, capillary retraction during the expansion, and the inertia of the bounding rim formed at the edge of the sheet. 
A simultaneous action of the restoring capillary force against the sheet surface formation and expansion, and the interplay of the liquid front with a less dense fluid (air), govern the deceleration and destabilization of the rim via a combined Rayleigh-Taylor-Rayleigh-Plateau (RT-RP) instability\,\myreplaced{\mbox{\citep{rayleigh1879capillary, taylor1950instability}} as investigated more recently, e.g., by \mbox{\citet{zhang2010wavelength, villermaux_drop_2011, javadi2013delayed}}; and \mbox{\citet{  wang_bourouiba_2018_rim}}}{\mbox{\citep{rayleigh1879capillary, taylor1950instability,  zhang2010wavelength, villermaux_drop_2011, javadi2013delayed,  wang_bourouiba_2018_rim}}}. 
The corrugations formed along the destabilizing rim can fragment further and shed microdroplets upon growth into ligaments. 
The rim is the critical connection between the evolving sheet and ligament shedding. 
In the particular case of fragmentation upon droplet impact on surfaces, \citet{wang_bourouiba_2018_rim} demonstrated that the rim thickness is the result of a dynamic adjustment balancing the volume influx from the sheet and the outflux from the end-pinching ligaments, while being subjected to time-varying inertial and capillary forces, the balance of which maintains a local and instantaneous rim Bond number Bo $=\rho (-\ddot{r}_{\textrm{s}}) b^2/\sigma=1$; thus, a rim thickness that adjusts to remain equal to the local and instantaneous capillary length $\ell_{\textrm{c}}\sim\sqrt{\sigma/\rho(-\ddot{r}_{\textrm{s}})}$, with $\ddot{r}_{\textrm{s}}$ being the rim deceleration  \myreplaced{\mbox{\citep{wang_bourouiba_2018_rim}}}{\mbox{\citep{wang_bourouiba_2017_thickness,wang_bourouiba_2018_rim}}}. 
Given this self-adjustment to maintain the local and instantaneous Bond number, Bo $=1$, the combined RT-RP fastest growing mode reduces to being close to that of the classical RP instability, despite the important effect of inertia and the RT contribution in setting the rim thickness balance. 
This powerful constraint, the \mbox{Bo $=1$} criterion, holds as long as sufficient liquid flows from the sheet into the rim to sustain ligament growth and their end-pinching, and thus maintains the We-independent relative mass distribution in the various parts of this system: sheet, rim, ligament, and droplets \myadded{as outlined in recent, connected works}\,\citep{wang_bourouiba_2021_growth,wang_bourouiba_energypartition,wang2023non} \myadded{that are directly applicable to the current study}.

Considering the crucial role of the nonlinear coupling of the rim with the continuous, unsteady, liquid volume influx from the sheet in shaping the rim's selection of number of ligaments and their droplet shedding, the absence of liquid influx could significantly alter the rim thickness self-adjustment, and the associated selection of number of ligaments, their growth and fragmentation. 
In this paper, we examine the following questions: 
 \begin{enumerate}
 \item Can we create the conditions of a severed rim to further clarify the role of fluid influx into a pre-formed rim?  In particular, we aim to determine the effect of interruption of such influx on the evolution of a freely evolving toroidal interfacial structure in radial expansion. 
\item  If so, how does the number of rim corrugations and ligaments evolve upon severance of the rim from the sheet, i.e., upon interruption of fluid influx from the sheet into the rim?
\item What is the final fragment number, or associated wavenumber,  from such a severed isolated rim in expansion? In particular, does it continue to follow the scaling laws inherited prior to severance or do spontaneous capillary re-configurations of the corrugations change the law, i.e., final fragment number?
\item On which timescale would a new rim re-from on a radially expanding sheet and how does it depend on the stage of the unsteady sheet expansion?
 \end{enumerate}

In answering the first question: we created the experimental conditions enabling the study of a severed rim from an expanding sheet, leveraging a laser-droplet interaction system in two steps: First, a tin microdroplet (diameters of $d_0$=30 and 40\,$\mu$m are considered) is subjected to a laser prepulse (energies in the range of $E_{\textrm{pp}}$=21--62\,mJ) enabling the spherical droplet to morph into a radially expanding sheet, spontaneously forming a rim. 
Second, the rim is detached by vaporizing the thin connecting neck with a second low-energy nanosecond laser pulse.
\myadded{We show that the vaporization pulse does not directly influence the fluid mechanics following rim severance. }
We analyze the rim pre and post severance to address the subsequent questions of interest above. 
We image the evolution of the system using stroboscopic imaging coupled with a double-framing camera (with a fixed delay of 650\,ns between frames). 
The advantage of our approach is to be able to time the severance of the rim at different stages of the sheet expansion. 
This is important to probe such unsteady system at various stages of its evolution. 
More on the methodology enabling to design this system is discussed in \S \ref{sec:exp} and the rim release is discussed in \S \ref{res:rimrelease}.  
 
Regarding the other questions, our analysis enables us to identify two distinct points of fragmentation of the severed expanding rim: 1) the base of the ligaments inherited from the pre-severance stage; and 2) the junction between inherited corrugations between such ligaments. 
We observe that the timescales of these two types of fragmentation points are not appreciably distinct. 
Our examination of the population of the ligaments and final rim fragments shows that the number of ligaments remains determined by the unsteady sheet expansion inherited from the attached rim prior to its severance. Moreover, the number of ligaments and corrugations continue to follow the theoretical predictions of  \citet{wang_bourouiba_2021_growth} when adjusted to the initial conditions of creation of the sheet, i.e., laser vs. surface-impact-induced impulse, and invoking the thickness profile  functional form of the droplet impact on surface of comparable size to that of the droplet, but adjusted to be consistent with the observed expansion trajectory\myadded{, without any further fitting parameters}. 
These findings are discussed in more detail in \S \ref{res:dynamics}\textcolor{red}{.1--4}. 
For the 4th question, we report that, post rim severance, the sheet re-forms a new bounding rim that continues to grow over time, itself re-forming new corrugations and ligaments, leading to final fragmentation. 
We discuss these final findings further in \S \ref{res:dynamics:innersheet}, with an argument to predict the onset of the rim re-formation.  
We highlight an interesting potential industrial application of mass re-distribution on the expanding sheet induced by such successive cycles of rim severance and re-formation in \S \ref{sec:appl}.


\section{Experimental method}\label{sec:exp}

\begin{figure}
\centering
\includegraphics[width=1\linewidth]{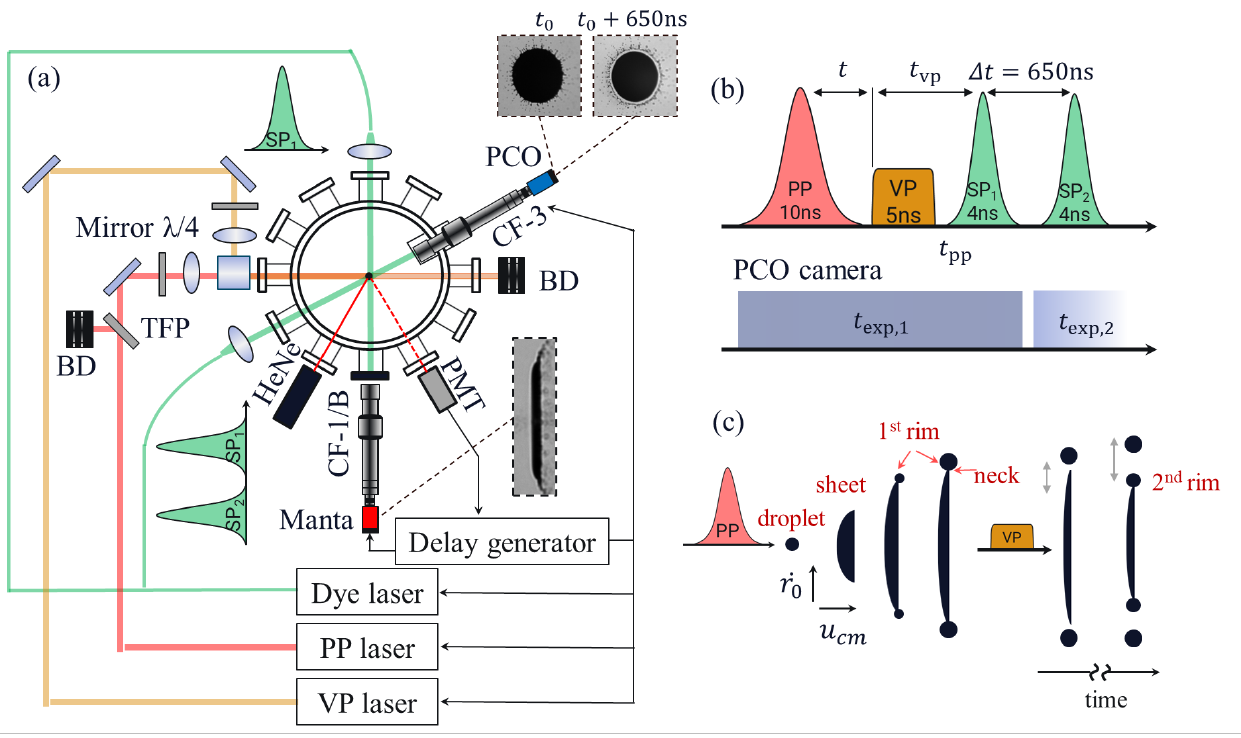}
\caption{(a) Top view of the experimental setup. Individual tin droplets are subjected to irradiation by multiple laser pulses. 
The synchronization system, triggered by the scattered light from a He-Ne laser, employs a delay generator to control the arrival time of all laser pulses. 
The energy of the pre-pulse (PP; seeded Nd:YAG Q-switched laser, Amplitude/Continuum Surelite) is tuned using a half-wave plate and a thin film-polarizer (TFP). 
A low-energy vaporization pulse (VP; Nd:YAG, quasi-continuous-wave diode-pumped \citep{Meijer:17}) is used to vaporize the tin, releasing the rim from the sheet. The remaining energy reflected from the TFP is blocked by a beam dump (BD).  
A stroboscopic imaging system is used to track the expansion of the tin sheet from the side (90\degree) and front (30\degree) with two probe pulses, $\textrm{SP}_\textrm{1}$ and $\textrm{SP}_\textrm{2}$ (see insets for the corresponding shadowgraphs). 
(b) Illustration of the sequences of laser pulses and camera exposures. 
The onset of the vaporization with VP is denoted by $t$. 
Subsequently, $\textrm{SP}_\textrm{1}$ is scanned over time at different moments after vaporization ($t_\textrm{vp}$). 
For the side view, only the first probe pulse is recorded whereas for the front view, a double-framing camera is used, where $\textrm{SP}_\textrm{1}$ falls within the first exposure time of the PCO camera $t_\textrm{exp,1}$ while $\textrm{SP}_\textrm{2}$ aligns with the second exposure time $t_\textrm{exp,2}$. 
See the main text for further details. 
(c) Conceptual sketch of the rim release. 
The formed rim is detached after VP impact. 
Later, the remaining sheet develops a new rim that grows over time.} 
\label{fig:1}
\end{figure}

The top view of the experimental setup used in this study is shown in fig.\,\ref{fig:1}(a). 
Further details of the setup can be found in prior work \myreplaced{\mbox{\,\citep[e.g.,][]{liu_speed_2022}}}{\mbox{\citep[e.g.,][]{liu_2020_mass, Liu_2021_PPRP, liu_speed_2022}}}. 
The laser-droplet interaction takes place in a high vacuum chamber that is kept at a $10^{-6}$\,mbar base pressure. 
We use a tank filled with solid tin placed on the top of the chamber. 
Tin is liquefied at $260\,\degree$C (well above its melting point of 232$\,\degree$C\,\citep{assael2010-tin-melting}) and the temperature is kept constant throughout the experiment.  
A pressure-driven droplet jet is generated through a micronozzle to produce two different droplet sizes, $d_0$=30 and 40\,$\mu$m, with density $\rho=7000\,\textrm{kg/m}^3$, surface tension $\sigma=0.544\,\textrm{N/m}$ and dynamic viscosity $\mu =1.8\times10^{-3}$\,Pa\,s \citep{assael2010-tin-melting}. 
The resulting droplet train is driven vertically downward at $\approx 10\,\textrm{m/s}$. 
Table \ref{table} summarizes the parameters used in these experiments, ranging from droplet sizes to prepulse laser energy, etc.   
When droplets cross a horizontal plane defined by the position on a He-Ne laser, which is directed a few millimeters above the center of the chamber, the scattered light is detected by a photomultiplier tube (PMT) attached to one of the ports. 
The resulting electric signal is down-converted to 10\,Hz and sent to a delay generator, which is used to trigger the rest of the laser and imaging systems. 
The data acquisition rate is set to the same frequency.  When a droplet passes through the defined horizontal plane, it interacts with a circularly polarized Gaussian beam with wavelength $\lambda_{\textrm{pp}}=1064\,\textrm{nm}$, and pulse length $\tau_{\textrm{pp}}=10\,\textrm{ns}$. 
The exact moment of irradiation is considered to be the instant when the peak intensity of the beam reaches the droplet surface. 
The laser pulse is focused onto a Gaussian spot on the droplet with a $\approx$102.5\,$\mu$m FWHM size at the focal point produced with a plano-convex lens (f=650\,mm). 
Before the laser-droplet interaction, the energy is tuned with a half-wave plate and a thin film polarizer.  To release the rim, we vaporize the sheet with a low-energy, spatially homogeneous (flat-top) laser pulse with 1064\,nm, 5\,ns pulse length and 822\,$\mu$m at FWHM (essentially its diameter); the energy of the laser pulse is individually tuned to a minimum value required to just detach the rim.   

To study the droplet dynamics after laser impact, we use a stroboscopic shadowgraphy imaging system based on incoherent, broadband light pulses with a wavelength of 564$\pm$10\,nm and 5\,ns pulse length at FWHM. 
These pulses are generated from the fluorescence of a Rhodamine 6G dye cell pumped by a 532\,nm laser pulse. 
The resulting light is guided into the chamber through optical fibers at 90\degree\ and 30\degree\ with respect to the prepulse propagation axis, to provide a side and (nearly) front view, respectively. 
Depending on the viewing angle, we use different camera configurations: For the side view, a single-framing CCD camera (Manta G145-B, AVT) is used with the CF-1/B objective (Edmund Optics), which captures a single probe pulse ($\textrm{SP}_\textrm{1}$). 
For the front view, we use a PCO-4000 double-framing camera, mounted with a CF-3 objective (Edmund Optics), that acquires two consecutive frames with a tuneable interframe delay. 
The effective interframe delay is determined by the time separation between two probe pulses ($\textrm{SP}_\textrm{1}$ and $\textrm{SP}_\textrm{2}$) and is set at 650\,ns. 
Thus, two independent shadowgraphs are obtained from the same sheet with a fixed time delay [see insets in fig.\,\ref{fig:1}(a)]. 
Figure\,\ref{fig:1}(b) displays a typical laser pulse sequence used in our experiments. 
The time elapsed after the pre-pulse (PP) and droplet interaction is denoted by $t$. 
A vaporization pulse (VP) hits the sheet at different set times.  
To capture the dynamics of the rim after vaporization, $\textrm{SP}_\textrm{1}$ reaches both cameras at $t_\textrm{vp}$. 
For the front view, a second $\textrm{SP}_\textrm{2}$ is captured after 650\,ns. 
Both $\textrm{SP}_\textrm{1}$ and $\textrm{SP}_\textrm{2}$ are captured by the camera (PCO) within exposure windows denoted as $t_\textrm{exp, 1}$ (explicitly $t_\textrm{exp, 1}=t_\textrm{vp}$) and $t_\textrm{exp, 2}$, respectively. 
The recording of the rim expansion starts a few nanoseconds before the vaporization of the neck. 
Note that the Manta camera uses only $\textrm{SP}_\textrm{1}$ within $t_\textrm{exp, 1}$, not shown in fig.\,\ref{fig:1}(b). 

\begin{table}
  \begin{center}
    \caption{Parameters presented in this study include droplet size $d_0$, capillary time defined as $\tau_{\textrm{c}}=\sqrt{\rho d^3_0/(6\sigma)}$, pre-pulse energy $E_{\textrm{pp}}$, initial velocity of radial expansion $\dot{r}_0$, deformation Weber number $\textrm{We}_\textrm{d}=\rho\dot{r}^2_0 d_0/\sigma$, deformation Reynolds number $\textrm{Re}_\textrm{d}=\rho\dot{r}_0 d_0/\mu$, with the dynamic viscosity of liquid tin $\mu = 1.8$ mPa$\cdot$s and its density  $\rho=7000\,\textrm{kg/m}^3$. 
    Note that the vaporization pulse laser energy range is 0.5--3\,mJ and is adjusted during the experiment until the first signs of rim detachment are visible. 
    } \label{table}
    \renewcommand{\arraystretch}{1.2} 
    \setlength{\tabcolsep}{10pt} 
    \small
    \begin{tabular}{lccccccc}
      \toprule 
      \toprule
      $d_0$ ($\mu$m) & $\tau_{\textrm{c}}$ ($\mu$s) & $E_{\textrm{pp}}$ (mJ) & $\dot{r}_0$ (m/s) & $\textrm{We}_\textrm{d}$ & $\textrm{Re}_\textrm{d}$ \\[3pt]
        \midrule 
       40  & 11.7 & 21 & 85 & 3702  & 13200 \\
       40  & 11.7 & 42 & 111 & 6358  & 17370 \\
       40  & 11.7 & 52 & 123 & 7808  & 19130 \\
       40  & 11.7 & 62 & 135 & 9364  & 21000 \\
       30  & 7.6  & 22 & 124 & 5914  & 14500 \\
       30  & 7.6  & 32 & 140 & 7476  & 16330 \\
       30  & 7.6  & 52 & 172 & 11250 & 20100 \\
      \bottomrule 
      \bottomrule
    \end{tabular}
  \end{center}
\end{table}

A conceptual illustration of the underlying physics discussed in this paper is presented in fig.\,\ref{fig:1}(c). 
Initially, the PP laser pulse hits a spherical droplet transforming it into a sheet that expands radially over time, bounded by a thick rim which is referred to as the "$1^\textrm{st}$ rim" situated at radial location $r_{\textrm{s}}$. 
Upon initial laser interaction, the droplet is propelled in the same direction as the laser at a constant center mass velocity $u_\textrm{cm}$ while simultaneously expanding radially at $\dot{r}_0$, \citep{kurilovich2016plasma,kurilovich2018power}. 
The ratio $\dot{r}_0/u_\textrm{cm}$ depends on both the laser energy,  the duration of the pulse and the droplet size \citep{HernandezRueda2022} and ranges $1.0 \lesssim \dot{r}_0/u_\textrm{cm} \lesssim 1.3$ in the current work.   
Subsequently, the VP vaporizes the thin neck connecting the sheet and the rim, allowing the latter to release and expand as well. 
Over time, due to the continuous outward radial flow of the expanding liquid sheet, an additional rim re-forms, which we refer to as the $2^{\textrm{nd}}$ rim, gradually increasing in size and eventually destabilizing into ligaments and rim fragments. 
According to our previous studies \citep{kurilovich2016plasma,kurilovich2018power,klein_drop_2020,liu_speed_2022}, the expansion dynamics of the laser-impulse-induced tin sheet is well captured by using a deformation Weber number $\textrm{We}_\textrm{d}=\rho \dot{r}_0^2 d_0 / \sigma$, based on initial radial expansion rate $\Dot{r}_0$ instead of the classical We related to the droplet impact speed on pillar $u_\textrm{cm}$ in \citet{wang2023non}. 
We maintain the parallel with the results of Wang and Bourouiba for impact on a pillar, differing in the initial impulse forming the sheet. 
However, given that $\Dot{r}_0=2u_0$ in the studies of \citet{wang2023non} -- due to initial impact conditions not fully understood -- we absorb a factor 4 ($\textrm{We}_\textrm{d}=4$We) in the We scalings of the prior works of these authors, wherever they appear in the remainder of this paper. 
In our experiments, we use a range of prepulse energies that result in different values for We and $\textrm{We}_\textrm{d}$, again noting that their ratio is not constant \citep{HernandezRueda2022}. 
Finally, the capillary time is defined as $\tau_{\textrm{c}}=\sqrt{\rho \mathbf{\textit{d}}_{0}^{3}/6\sigma}$ and for droplet sizes $d_0$=30 and 40\,$\mu$m it equals 7.6 and 11.7\,$\mu$s, respectively (see Table \ref{table}).

\section{Rim release}\label{res:rimrelease}
\begin{figure}
\centering
\includegraphics[width=1\linewidth]{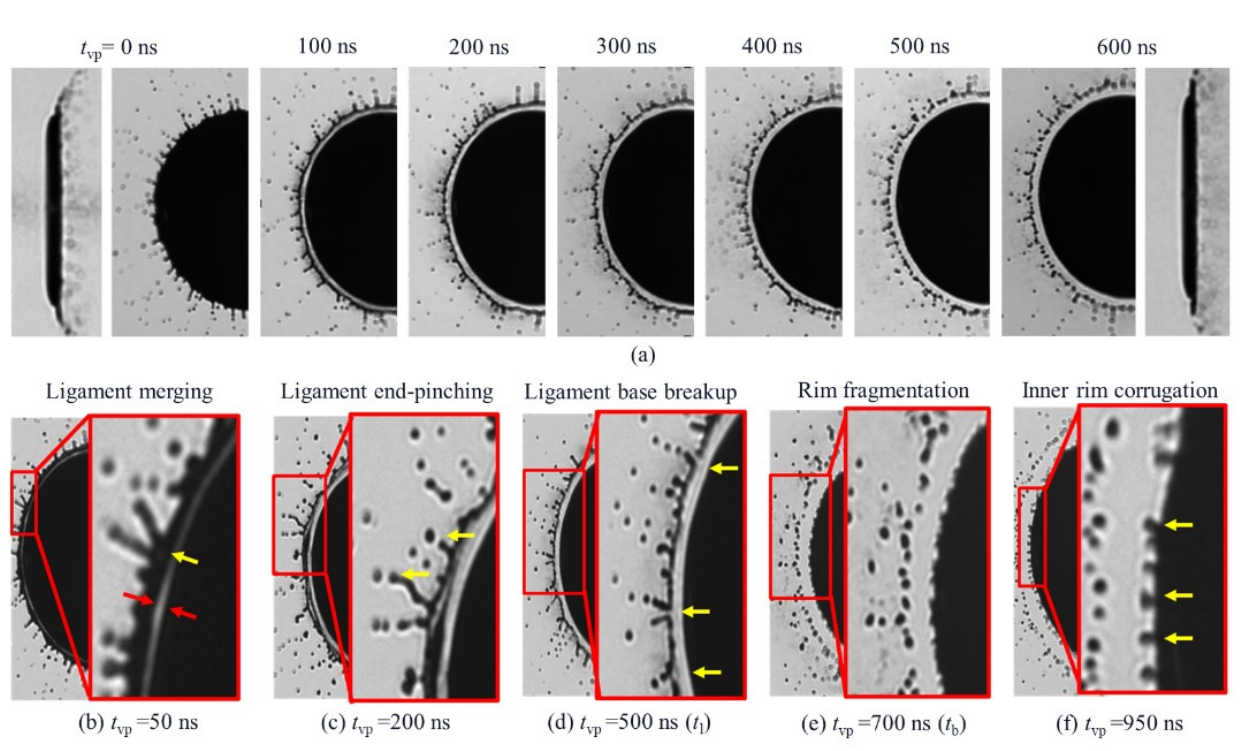}
\caption{Phenomenological description of rim severance and subsequent destabilization.
(a) Front-view images depict the sheet before ($t_{\textrm{vp}}$=0\,ns) and after ($t$=100--600\,ns) interaction with the VP laser. 
\myadded{Two side-view images (for $t_{\textrm{vp}}$=0 and 600\,ns) are additionally shown. }
Shadowgraphs were captured using a droplet with an initial size of $d_{0}$=40\,$\mu$m, PP laser energy of $E_{\textrm{pp}}$=21\,mJ, leading to $\textrm{We}_\textrm{d}$=3702, and using a VP laser energy $E_{\textrm{pp}}$=3.2\,mJ. 
Recall that the VP laser pulse energy is adjusted independently for each droplet size and Weber number to achieve, just, a clear rim severance. 
Quantification of breaking is done on the left-hand side of the sheet, since the front view shadowgraphs are recorded with $30\degree$\ resulting in a partially out-of-focus view of fragments on one side of the sheet. 
(b--f) Illustration of the sequence of hydrodynamic destabilization following rim detachment, progressing in time from left to right. 
(b) VP laser impact \myadded{detaches the rim already at the end of the VP and } initiates a ballistic expansion of the rim, manifested as a translucent gap (red arrows), observable some 50\,ns after VP impact. 
Previously formed ligaments exhibit base merging (yellow arrows). 
(c) By 200\,ns, a typical event of end-pinching fragmentation of the ligament occurs.
\myaddedred{The finite (spatial) coherence of the backlighting is the origin of minor diffraction effects producing slightly brighter regions near sharp or small features.} 
(d and e) The breakup of the rim at two distinct points: the base of the ligaments at $t_l$=500\,ns, and the rest of the corrugated rim at $t_b$=700\,ns. 
(f) Around $t$=950\,ns, inner sheet corrugation and further formation of ligaments are observed.}
\label{fig:2}
\end{figure}

Figure \ref{fig:2}(a) shows several frames to illustrate some of the hydrodynamic behavior of the rim post severance. 
\myadded{Side-view images (for $t_{\textrm{vp}}$=0 and 600\,ns) are additionally shown and illustrate the finite curvature of the sheet, as described by \citet{Franca2025}, which has negligible influence on the measurement of the dominantly radial expansion of the rim and ligaments. }
The VP laser pulse starts the vaporization of the resulting sheet 2$\,\mu$s after PP impact, well before $t_\textrm{max}$, the time of maximum radial extension of the sheet attained at $t_\textrm{max}=0.38\tau_{\textrm{c}}\approx 4\,\mu s$ \citep{villermaux_drop_2011,klein_drop_2020}. 
The VP laser pulse energy is then adjusted to just achieve clear rim severance. 
According to \citet{klein_drop_2020}, the neck binding the rim to the sheet can be considerably thinner than the rest of the sheet. \mydeleted{ and only a negligible (if any) fraction of the rim is vaporized.} 
The underlying vaporization mechanism is detailed by \citet{Schubert2024a}. 
We note that the VP intensities are of the order $\sim 10^{7}$\,W\,cm$^{-2}$\myadded{, up to $9 \times 10^{7}$\,W\,cm$^{-2}$ for the highest energy case, } well below the threshold of producing any plasma [$\sim$$\myadded{4 \times}  10^{8}$\,W\,cm$^{-2}$ under the current conditions \citep{Engels2025}] or any propulsion \citep{kurilovich2016plasma} such as associated with the PP (here $\sim 10^{10}$\,W\,cm$^{-2}$). 
\myadded{In sum, no plasma or related plasma propulsion occurs in our experiments due to the VP.}
\myadded{Next, we estimate the amount of vaporization of the sheet (including the thin neck connecting the sheet to the rim) on the one hand, and the thick rim on the other.
For the thin film sheet ($\sim$10--100\,nm, in the thin-film limit of \mbox{\,\citet{Schubert2024a}}), we use the robust relation between the vaporization rate and the laser fluence, as established by \mbox{\,\citet{Schubert2024a}} and estimate that the thickness vaporized from the sheet before the rim clearly separates to be $2-11\,$nm, depending on the set intensity of the VP. }
\myadded{For the much thicker bulk ($\sim$$\mu$m) rim we cannot use the aforementioned scaling relation (being valid in the thin-film limit) and instead apply the full numerical model of \mbox{\,\citet{Schubert2024a}}. 
The surface temperature predicted by the model remains below 1800\,K, even for the highest energy case, given the fast diffusion of heat into the bulk. 
At these low temperatures, no appreciable vaporization occurs on the rim (tin boils at 2875\,K). 
Furthermore, the bulk liquid rapidly distributes heat and after 50\,ns a heating of only 200\,K is induced overall, leading to a negligible, local 5\% reduction in surface tension accompanied by a 2\% reduction in density. 
Thus, whereas the VP leads to modest vaporization of the thin-film sheet and fully removes the thin neck connecting the rim, the much larger bulk of the rim does not vaporize or significantly heat. }

Figures \ref{fig:2}(b-f) show the evolution of the rim after detachment. 
Once the neck is removed, the rim rapidly separates from the sheet, given the ballistic trajectory of the rim (see \mydeletedred{Appendix\,\ref{A.2} and }below) and the retraction of the sheet edge to form a new rim (see below). 
The physical separation is clearly visible in fig.\,\ref{fig:2}(b), represented by red arrows, as early as $t_\textrm{vp}$=50\,ns. 
As the visible gap increases over time, we note that the ligaments stop growing, as expected in the absence of liquid input from the sheet, and they appear as static liquid protuberances that continue to fragment by capillary tension, consistent with ligament end-pinching and RP instability of the toroidal severed rim \citep{Rayleigh_instability_jet, rayleigh1879capillary} and as shown in fig. \ref{fig:2}(c) \citep{castrejon2015plethora, wang_bourouiba_2021_growth}. 
After the separation of the rim from the sheet, the rim rapidly starts to fragment. 
We identify two parts of the rim with significant curvature that are prone to pinching: the base of the ligaments and the corrugated parts of the main rim. 
Indeed, the rim is seen to break first at these two types of locations, cf. fig.\,\ref{fig:2}(d,e). 
Finally, over time, the inner sheet forms a second rim. This \emph{rim recurrence} is first visible as a corrugation in fig.\,\ref{fig:2}(e) from which a myriad of ligaments start to emerge at a later time [see fig.\,\ref{fig:2}(f)]. 
Further growth of these ligaments leads to the shedding of droplets (see \S \ref{res:dynamics:innersheet}). 
In the following sections, we proceed to quantify the features of rim destabilization, including the relevant timescales of capillary-driven breakup, and the resulting selection of the final population of ligaments and fragments.

\section{Dynamics of the severed rim}\label{res:dynamics}
\subsection{Sheet expansion}\label{res:dynamics:sheet}
\label{IV.1}

\begin{figure}
\centering
\includegraphics[width=1\linewidth]{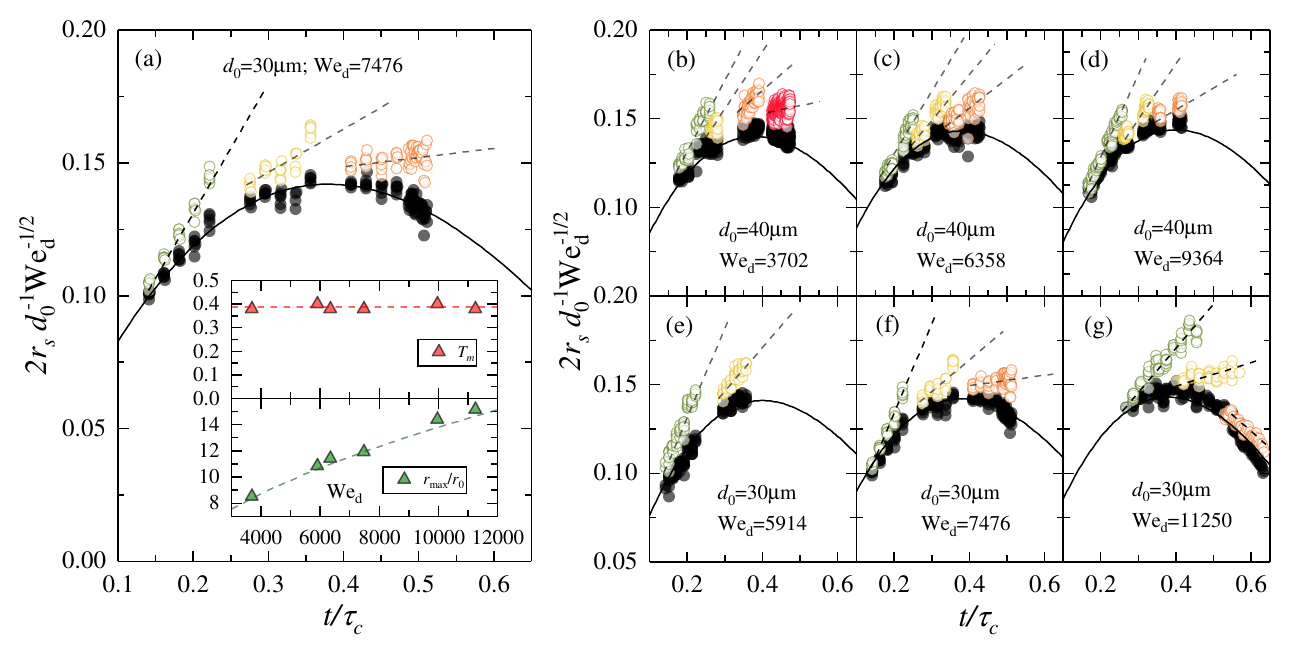}
\caption{Sheet and rim expansion after detachment. 
Two different droplet sizes were used: $d_0$=30 and 40\,$\mu$m, with a range of $\textrm{We}_\textrm{d}$ values. 
Solid black points correspond to the evolution of the sheet after rim release, while open colored points correspond to the rim diameter after having been detached at different moments of the sheet expansion.  Each $d_0$-$\textrm{We}_\textrm{d}$ combination is displayed within the same non-dimensional time range, where the release of the rim takes place at different times. 
We consider $2r_{\textrm{s}} d_0^{-1}\textrm{We}_\textrm{d}^{-1/2}$ as a non-dimensional radius. 
(a) A representative case of an expanding sheet and rim formed from a droplet with an initial diameter of $d_0=30\,\mu$m hit by PP laser with energy $E_{\textrm{pp}}$=32\,mJ to radially expand with $\textrm{We}_\textrm{d}$=7476. 
\myadded{Linear fits capturing the ballistic trajectory followed by the rim over time are indicated with dashed lines. The solid line shows the evolution captured by (\ref{Eq:1})} \mydeleted{using the fitted parameters (see the main text)}.   
The diameter of the rim is measured within $0 - 1.4\,\mu$s after the vaporization, a range that changes in non-dimensional time units for each droplet size.
The inset in (a) shows the consistency of $r_{\textrm{max}}/r_0$ and $T_{\textrm{m}}$ for different $\textrm{We}_\textrm{d}$. The red dashed line illustrates the constant value of $T_{\textrm{m}}=0.38$ and the green dashed line depicts $r_{\textrm{max}}/r_0\sim0.14\textrm{We}^{1/2}_\textrm{d}$ (see  \S.\,\ref{IV.1} for details). 
(b--g) Expansion curves of all droplet sizes and $\textrm{We}_\textrm{d}$ studied during the experiments. }
\label{fig:3}
\end{figure}

Figure \ref{fig:3} shows the position of the outer edge of the sheet and rim, where the rim is vaporized at different instances of sheet expansion. 
For a detailed view,  fig.\,\ref{fig:3}(a) shows the particular case in which the rim is released at times $t$=1, 2, and 3$\,\mu$s after PP laser impact, with VP laser energies $E_{\textrm{vp}} \approx$ 1.0, 1.0, and 0.8\,mJ, respectively. The non-dimensional radius scaled with deformation Weber number $ r_{\textrm{s}} r_0^{-1}\textrm{We}_\textrm{d}^{-1/2}$  is shown evolving over non-dimensional time $t/\tau_{\textrm{c}}$. \myadded{Here, $r_{\textrm{s}}$ is the sheet radius with  $r_0=d_0/2$}. We observe two different trajectories, one corresponding to the expansion of the freely expanding severed rim and the other to the remaining sheet.

To rationalize the expansion of the sheet, recall the third-order polynomial equation derived analytically by \citet{wang2023non} to describe the expansion  of the unsteady sheet produced from the impact of a water droplet on a pole of comparable size to that of the impacting droplet: 
\begin{equation}
    2 R_{\textrm{s}} \equiv 2 r_{\textrm{s}}\mathbf{\textit{d}}_{0}^{-1}=\textrm{We}_\textrm{d}^{1/2}(b_3(T-T_{\textrm{m}})^3+b_2(T-T_{\textrm{m}})^2+b_0),
\label{Eq:1}
\end{equation}
with $T=t/\tau_{\textrm{c}}$, $T_{\textrm{m}}$ as the non-dimensional (apex) time of maximum sheet expansion and $b_0$ the corresponding non-dimensionalized maximum sheet radius scaled as $b_0=r_{\textrm{max}}/(r_0\sqrt{\textrm{We}_{d}})$. 
The rest of the coefficients have the following physical meaning\myadded{, as detailed by \citet{wang2023non}}: $2b_2$ is the deceleration of the rim velocity along the sheet expansion, while $b_3$ is related to the initial velocity of expansion. 
Prior studies indicate that the tin sheet's evolution is better captured when considering the deformation Weber number, $\textrm{We}_\textrm{d}$, based on the initial radial expansion rate $\dot{r}_0$ instead of the $\textrm{We}$ related to the center mass translation velocity $u_\textrm{cm}$\,\citep{liu_speed_2022}. 
Indeed, this may reflect the difference in initial impulse inducing the transfer of vertical to horizontal momentum upon the sheet's early formation: On the one hand,  a laser-driven liquid tin sheet forms from impulse generated by the pressure field imprinted by the plasma produced from the droplet surface; on the other hand, a sheet is formed from surface boundary impact with the associated pressure field.  
Given that $\dot{r}_0\approx 2 u_\textrm{cm}$ was reported by Wang and Bourouiba \citep{wang_bourouiba_2021_growth,wang2023non}, we make the correspondence to their work by absorbing a factor 4 ($\textrm{We} \propto u^2$, so $\textrm{We}_\textrm{d}=4\textrm{We}$ in their work), consistent with the previous formalism in \citet{liu_speed_2022}. 
\myadded{In the present study, the measured time-evolution of the tin sheet radius  is consistent with  $T_{\textrm{m}}=0.38(1)$ and $b_0=0.14(1)$, $b_2=0.58(2)$ and $b_3=0.43(4)$, similar to prior works on tin sheets \citep[cf.][]{Liu_Bo_phdthesis}. We adopt these coefficients for the sheet dynamics evolution and all related calculations in all subsequent parts of this manuscript, including the discussion of the rim dynamics \S\,\ref{IV.11}, the ligament base breakup and rim fragmentation \S\,\ref{IV.2}, the number ligaments and fragments \S\,\ref{IV.3}, and the inner sheet behavior \S\,\ref{res:dynamics:innersheet}.  
Note that we shall discuss in more detail the coefficients and the physical interpretation of the comparison between the current laser-impact-droplet  and the water-pole impact systems Appendix \S \ref{A.3}.}

\mydeleted{Recall that \mbox{\citet{wang_bourouiba_2017_thickness} derive} the coefficients $\left\{b_0,b_2,b_3 \right\}$ from the experimentally validated thickness profile of the expanding sheet (see Appendix \S \ref{A.3}\textcolor{red}{.1}) combined with the instantaneous and local rim Bo $= 1$ condition governing the rim's thickness \mbox{\citep{wang_bourouiba_2018_rim,wang2023non}}.  Recalling \mbox{\citet{wang_bourouiba_2021_growth,wang2023non}}, here we define the non-dimensional inertial time as $T^* =  \dot{r}_0 t/2 d_0$,  the radial coordinate as $R = r/d_0$, and the thickness as $H\equiv h/d_0$.}
\mydeleted{The authors show that the thickness profile of the  radially expanding water sheet from the impact on a pole of comparable size to that of the impacting droplet is self-similar and captured by $F_{\textrm{wp}} (x)=1/(a_1X+a_2X^2+a_3X^3)$ in the non-dimensional thickness profile $H=F/T^{*2}$, with similarity variable $X=R/T^*$ and $\left\{a_1=24.4(2), a_2=-38.1(4), a_3=35.2(3) \right\}_{\textrm{wp}}$, and where the numbers in brackets indicate standard deviation. Here, "$\textrm{wp}$" is used to denote "water sheet expanding from droplet impact on pole". The set of thickness profile coefficients $\left\{a_{\textrm{i}} \right\}_{\textrm{wp}}$  self-consistently leads to expansion coefficients $\left\{b_{\textrm{i}} \right\}_{\textrm{wp}}$ with  reasonable agreement with the laser-droplet case cf. \S \ref{A.3}\textcolor{red}{.1}. }

\mydeleted{However, the initial condition of impulse creating the radially expanding thin sheet differ between the water droplet impact on pole and the tin drop laser impact. Hence, although the expanding sheet dynamics for the laser impact on tin droplet, which we measure here, is consistent  with the functional form of \mydeleted{\citet{wang_bourouiba_2017_thickness}}, the coefficients of the profile differ. This difference is inherently related to the initial impact timescale and early sheet formation. There remains a need to clarify theoretically how such early sheet formation dynamics prescribes these coefficients for the laser impact on the tin droplet system.  In the meantime, in this work, we adopt the  validated functional form of the water-impact thickness profile of \mydeleted{\citet{wang_bourouiba_2017_thickness}} with adaptation of its constants to reflect the differences in initial impulse condition. This approach leads to $\left\{a_1=28.6(2), a_2=-26.2(1), a_3=23.5(3) \right\}_{\textrm{tw}}$ using  $\left\{b_{\textrm{i}} \right\}_{\textrm{tw}}$ coefficients as input from the measured  sheet radial expansion curve. We denote this profile and these coefficients with "$\textrm{tw}$" for "this work". }

\mydeleted{We note that other thickness profiles were proposed in the literature for laser impact on tin droplet, for example with heuristically modified  functional form  in \mbox{\citet{liu_2020_mass}}  $F_{\textrm{ef}}(x)=1/(a_0+a_1X+a_2X^2)$, with $\left\{a_0=1.65(2), a_1=6.9(3), a_2=-2.4(8) \right\}_{\textrm{ef}}$ obtained from thickness measurement of  radially expanding tin sheets. We denote this profile with  "$\textrm{ef}$" standing for "empirical fit".  In the present study, the tin sheet experimental data  shows  $T_{\textrm{m}}=0.38$ and $b_0=0.14$, $b_2=0.58$ and $b_3=0.43$ from a global fit  \mbox{\citep[e.g.,][]{Liu_Bo_phdthesis}}. The constraint on the sheet expansion prescribed by these $\left\{b_0,b_2,b_3 \right\}$ coefficients leads, in turn, to coefficients for the empirical thickness profile that are not in agreement with the experimental tin sheet thickness profile measurements in  \mbox{\citet{liu_2020_mass}} (see Appendix\,\ref{A.3}). }

\mydeleted{In sum, for the present work, we remain consistent with the validated theoretical results of \mbox{\citet{wang_bourouiba_2017_thickness,wang_bourouiba_2021_growth,wang2023non}} and adopt the $F_{\textrm{wp}} (x)$ functional form for the tin sheet thickness profile; however, we also account for the distinction in initial impulse by accounting for the difference in thickness profile constants from $\left\{a_{\textrm{i}} \right\}_{\textrm{wp}}$ to $\left\{a_{\textrm{i}} \right\}_{\textrm{tw}}$ so that the $\left\{b_0,b_2,b_3 \right\}$ of the tin sheet expansion remain consistent with the current measurements. In the remainder of this work, "$\textrm{tw}$" denotes "this work" and indicates that the functional form of the thickness profile of \mbox{\citet{wang2023non}} is used with adjusted constants $\left\{a_{\textrm{i}} \right\}_{\textrm{tw}}$. }

\mydeleted{The coefficients of the tin radially expanding sheet  compare with those of the theoretically derived and validated water sheet expansion in the air upon impact as follows: } 
\iffinalmode
\mydeleted{
\else
\begin{itemize} 
\item \mydeleted{The maximum sheet expansion time for our laser-on-tin experiments is $T_{\textrm{m}}=0.38$ while it is $T_{\textrm{m,p}}\approx 0.43$ for the water droplet impact on a pole \mbox{\citep{wang2023non}}, i.e., $T_{\textrm{m}}/T_{\textrm{m,p}}\approx0.9$. 
This shift to earlier sheet radius apex time could be attributed in part to mass ablation from the droplet surface upon laser impact. 
In fact, previous studies estimated  approximately $10-20\%$ mass loss from ablation \mbox{\citep{HernandezRueda2022,liu_2023_mass}}. 
Such mass ablation would result in a shifted capillary time:  $0.95-0.9\tau_{\textrm{c}}$.  Furthermore, changes in liquid temperature could affect surface tension and density. However, their ratio, relevant for hydrodynamic equations, suggest a weak dependence on temperature \mbox{\citep{Liu_Bo_phdthesis}}.}

\item\mydeleted{We observe a ratio $b_0/b_{\textrm{0,p}}\approx 1.2$ for the coefficient $b_0$ associated with the maximum sheet radius. The differences in both $T_\textrm{m}$ and $b_0$ values are consistent also with a reduced shear stress, i.e.,  energy loss, from a surface-free laser-driven sheet expansion expected to result in a larger maximum sheet radius. Moreover, relative to the  impact on pole, the "over-expansion" of the tin sheet is reminiscent of higher  maximum sheet expansion of \mbox{\citet{lastakowski2014bridging}} for impact on superheated solid surfaces  compared to impact on pole \mbox{\citep[e.g., figure 16 in][]{wang2023non}}.}

\item\mydeleted{Regarding $b_2$, associated with the deceleration of the rim: the ratio $b_2/b_{\textrm{2,p}}\approx 1.4$, indicates a larger decrease in sheet velocity for the laser-driven expansion compared to that which is pole-impact driven. }

\item\mydeleted{Finally, the ratio $b_3/b_{\textrm{3,p}}\approx 2.6$ reflects a  larger initial expansion rate for laser-driven tin sheets. }
\end{itemize}
\iffinalmode
}
\else
\fi
\mydeleted{Overall, these distinctions are further reminders of the fundamental differences in  initial impulse conditions causative of the radially expanding sheet formation in the two systems compared here. 
For example, the presence of a bulky central mass after sheet contraction in liquid tin \mbox{\citep{Liu_Bo_phdthesis}} could be consistent with  possible cavitation inside the droplet, hence, compressibility effects on timescales $\sim\tau_{\textrm{pp}}$ \mbox{\citep{reijers2017,liu_2023_mass}}. 
Taken together, these observations further reinforce the need for future theoretical and experimental investigations on  how the initial impulse causative of the radially expanding  sheet formation differ in the laser-on-tin impact versus water droplet impact on a pole of comparable size to that of the impacting drop.}

\subsection{Rim dynamics}
\label{IV.11}
Figure \ref{fig:3} shows that the detached rim follows a ballistic trajectory.  Indeed, the dashed lines reflect linear fits to the rim position over time \myadded{from the moment the rim has visibly separated from the sheet}. 
Note the intersection of these lines with the expansion curve at the moment of rim severance. 
As will be discussed in more detail hereafter, this finding supports the idea that the expansion rate of the rim is determined by the local instantaneous velocity $\dot{r}_{\textrm{s}}$ of the sheet rim at the time of severance. Given the $\approx 5\,\mu$m resolution of the imaging system, the sheet can be traced independently starting some 100\,ns after rim release.
\myadded{The release of the rim itself occurs earlier, on the 5-ns timescale of the VP. } 
On the remaining sheet, the continuously outward-flowing liquid immediately starts \myadded{to form} \mydeleted{from} a new rim bounding the sheet. 
The thickness of this second rim increases with time and becomes of the same order as the local capillary length, when the unsteady evolution of the sheet's rim satisfies the criterion of local, instantaneous Bo $=1$\,\citep{wang_bourouiba_2018_rim}. 
However, the rim requires a transition time to satisfy this criterion. 
We estimate this transition time  to range from $\approx1$\,$\mu$s at early rim severance to a few 100\,ns at late rim severance times (see \S\,\ref{res:dynamics:innersheet} for more details). 

Given these estimations, we can expect a time-varying response of the inner sheet immediately post rim severance. 
The sheet is expected to follow a ballistic motion before the velocity is reduced by capillary forces, enabling the re-formation of the second rim which eventually grows sufficiently to be subject to the local force balance imposing the universal rim criterion Bo $=1$ again. 
Beyond this effect, the sheet appears to follow a trajectory indistinguishable from the initial pre-severance one. 
Regarding the severed free toroidal rim, it is set in motion immediately post-severance, with ballistic expansion up to complete and final fragmentation. 
As we shall see in the next section, the rim undergoes a rapid fragmentation over $\approx 100-500$\,ns. Considering the rim as a perfect cylindrical jet with average thickness $\approx5\,\mu$m, the corresponding capillary time is on the order of several\,$\mu$s, which is much longer than the time required for the complete fragmentation of the rim, thus weakening any visible capillary-driven retraction of the rim. 
\mydeletedred{Furthermore, as we discuss in Appendix\,\ref{A.2}, the vaporization of the thin connecting neck imprints a negligible momentum on the rim (or sheet), so no significant acceleration from vapor expansion is expected or indeed observed \mydeletedred{in \S\, Appendix\,\ref{A.2}}. 
Given these arguments, we expect the ballistic expansion to be determined solely by $\dot{r}_{\textrm{s}}$ at the detachment/severance time. 
Given the sheet deceleration, a late-time detachment considerably reduces the expansion rate of the rim upon severance, with a retraction for severance occurring post sheet expansion apex, as seen in fig.\,\ref{fig:3}(g).}

\begin{figure}
\centering
\includegraphics[width=1\linewidth]{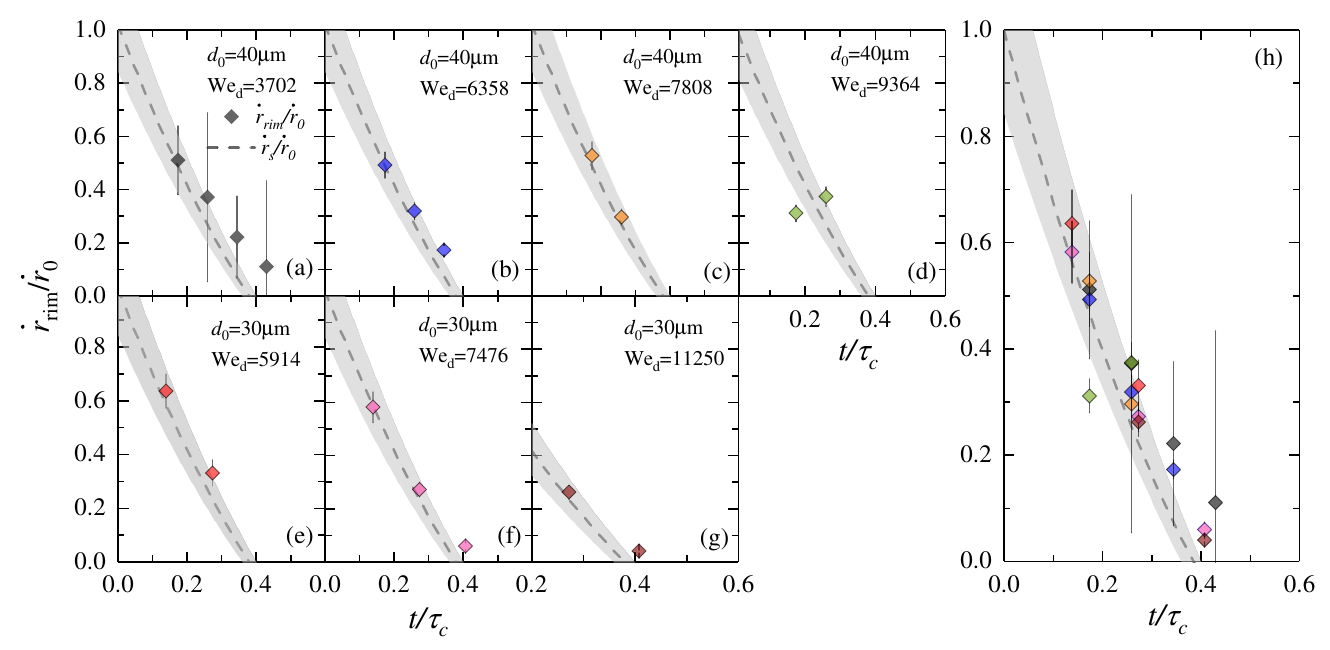}
\caption{\myaddedred{(a--g) Rim velocity, $\Dot{r}_\textrm{rim}$, after detachment scaled with the initial velocity of the expanding sheet, $\dot{r}_0=\textrm{We}_\textrm{d}d_0/\tau_{\textrm{c}}$, for several $d_0$ and $\textrm{We}_\textrm{d}$, as a function of the non-dimensional time, $t/\tau_{\textrm{c}}$. 
(h) All represented data are displayed collectively. 
The dashed line corresponds to the instantaneous sheet edge velocity, $\dot{r}_{\textrm{s}}$, scaled with $\dot{r}_0$. The gray shaded area indicates the one-standard deviation uncertainty of the velocity as derived from the uncertainties in the coefficients $b_2,b_3$ and $T_m$. }}
\label{fig:A2}
\end{figure}

\myaddedred{A detailed analysis of the rim velocity after severance shows a ballistic motion of the rim, as seen in fig.\,\ref{fig:3}. 
We find that the rim maintains a constant speed,  approximating the speed of the rim at the moment of its severance from the sheet cf. fig.\,\ref{fig:A2}(a--g).
No significant difference is found between the velocity of the rim and the velocity of the sheet prior to rim severance [see fig.\,\ref{fig:A2}(h)].
The vaporization of the thin connecting neck imprints a negligible momentum on the rim (or the sheet), underpinning the observation of no acceleration.
Suboptimal experimental conditions related to imperfections in the generation of the droplet stream for the $d_0$=40\,$\mu$m, We$_d$=3702 case in panel (a) leads to larger scatter in the data, reflected by corresponding larger uncertainties shown in the figure. 
}

\subsection{Ligament base breakup and rim fragmentation}
\label{IV.2}

\begin{figure}
\centering
\includegraphics[width=1\linewidth]{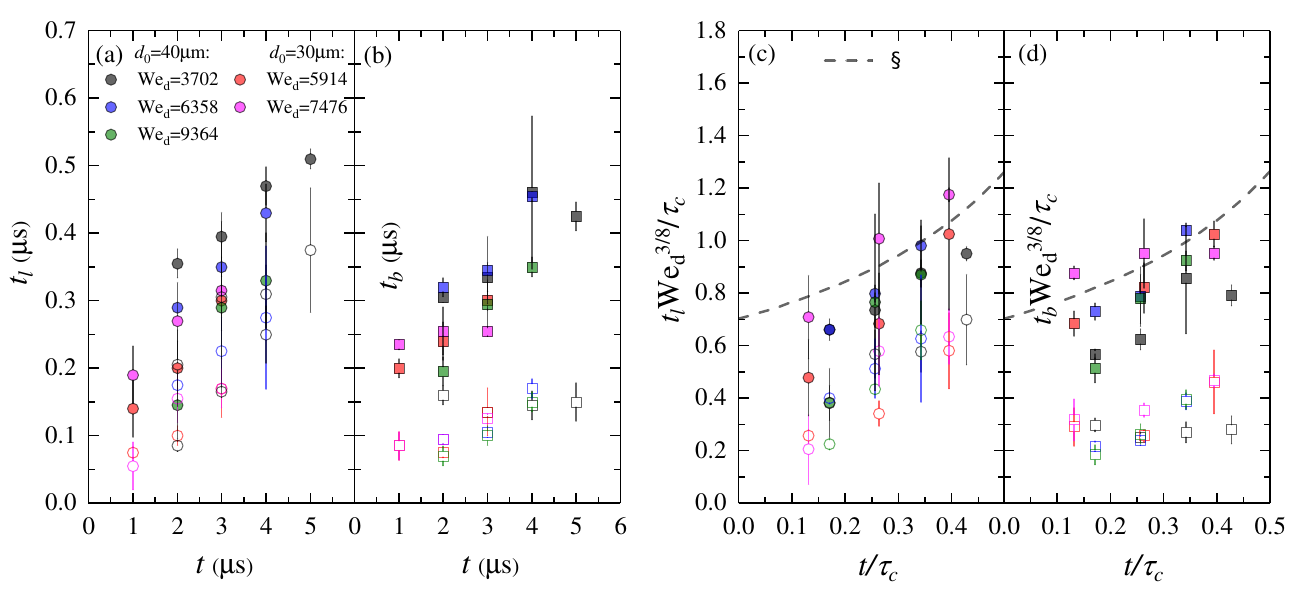}
\caption{Ligament base and rim breakup times. 
(a) Ligament base breakup times $t_\textrm{l}$ for different times of rim severance. 
The first and last breaking instances are denoted as open and solid points, respectively. 
Each color corresponds to a different pair of $d_0$ and $\textrm{We}_\textrm{d}$. 
(b) Rim fragmentation time $t_\textrm{b}$, for the same dataset. 
(c and d) Same data for $t_l$ and $t_b$ scaled as $t_\textrm{l,b}\textrm{We}_{\textrm{d}}^{3/8}/\tau_{\textrm{c}}$ as a function of non-dimensionalized time, $t/\tau_{\textrm{c}}$. 
The dashed lines corresponds to (\ref{Eq:2}) \myadded{, which should be compared to the solid markers. No fitting is performed.}  }
\label{fig:4}
\end{figure}

The flow of liquid from the sheet significantly shapes the behavior of the thickness of the rim, the subsequent growth of ligaments, and their final fragmentation \citep{wang_bourouiba_energypartition}. Once the rim is severed from the sheet, the inflow of liquid ceases, while the rim continues to expand radially.  
As already mentioned, in our experiments, we observe that the rim breaks at two (families of) locations: at the base of the ligaments (at time $t_l$) and along the rest of the rim between adjacent ligaments (at time $t_b$). 
Leveraging images similar to fig. \ref{fig:2}(a), we extract the breaking times shown in figs. \ref{fig:4}(a) and \ref{fig:4}(b) for all datasets used in this study. 

We consider two breaking times: the time of observation of the very first breakup and the time at which  full fragmentation of the rim  completes. 
For each studied detachment time $t$, approximately 20 frames are recorded. 
Our analysis splits up these frames into two groups of 10 frames each. 
Each group of frames results in a value for early and late breaking times, and we report on the values averaged over the two sets with the uncertainty defined as the standard deviation (difference) between the sets of frames. 

Figures \ref{fig:4}(a) and (b) show that the breaking time for both the rim-ligament junctions and the rim increases monotonically with time. 
This increase can be understood by combining the \mbox{Bo $=1$} criterion with the characteristic RP-breakup time of a cylindrical jet $ t_{\textrm{br}}\approx 2.91\sqrt{\rho b^3/8\sigma}$ \citep{rayleigh1879capillary} for a rim of diameter $b$ or equivalently, for a ligament base with width $w$, taking  $w(t) \approx b(t)$ \myadded{following, e.g.,} \citet{wang_bourouiba_2021_growth}. 
Thus, the characteristic time  $t_\textrm{l}\sim w^{3/2}$ and $t_\textrm{b}\sim b^{3/2}$ both grow with  rim thickness. 
\myadded{The prefactor 2.91 is obtained from the reciprocal of the growth rate (0.344) of the fastest growing RP mode following \mbox{\citet{rayleigh1879capillary}}, as supported by direct measurements of necking times by \mbox{\citet{clanet_transition_1999}}.}

\myadded{We next obtain $b(t)$ from the Bo $=1$ condition as $b(t)=[\sigma/(-\ddot{r}_s\rho)]^{1/2}$, where $\ddot{r}_s=d_0\sqrt{\mathrm{We_d}}/2\tau_\mathrm{c}^2\left[2b_2+6b_3\left(T-T_\mathrm{m}\right) \right]$ is the sheet acceleration derived from (\ref{Eq:1}). Knowing the expansion coefficients ${b_2,b_3}$ and combining $b(t)$ with $t_\mathrm{br}$, the corresponding rim fragmentation time, after some algebraic manipulation, is stated as}:


\begin{equation}
    t_{\textrm{br}}/\tau_{\textrm{c}}=\myadded{1.1\textrm{We}_\textrm{d}^{-3/8}[6b_3\left(T_\mathrm{m}-T \right)-2b_2]^{-3/4}=}1.1\textrm{We}_\textrm{d}^{-3/8}(2.1-2.4T)^{-3/4},
\label{Eq:2}
\end{equation}
\myadded{with the prefactor 1.1 originating from the combination of constants in $t_{\textrm{br}}, \tau_{\textrm{c}}, b, \ddot{r}$, resulting in $2.91 \cdot (6/8)^{1/2} \cdot 6^{-3/4} \cdot 4^{3/8}\approx 1.1$.
We justify invoking the Bo $=1$ condition by considering the preconditions of a local Reynolds number \mbox{$\textrm{Re}=b v_b/\nu \gtrsim 8$} \citep{wang_bourouiba_2018_rim},
with characteristic velocity $v_b=\sqrt{\sigma/\rho b}$.  
In our experiments, the corresponding local Reynolds number ranges $30<\textrm{Re}<70$, fulfilling the specified preconditions.  
We are not aware of an elastic component to tin, so here we omit the  condition related to viscoelasticity as described by \citet{wang_bourouiba_2018_rim}. 
}

Figures\,\ref{fig:4}(c) and \ref{fig:4}(d) show the non-dimensional \myreplaced{breakup}{breaking} time scaled with $\textrm{We}_\textrm{d}$ over non-dimensional time $t/\tau_{\textrm{c}}$, both for the first and last instances of fragmentation. 
\myadded{Although the spread in the data is not significantly reduced, the rescaling enables the comparison to the theoretical prediction. }
Equation (\ref{Eq:2}) captures the timescale at which \emph{full} fragmentation occurs well. 
However, the first instances of rim and ligament breakup occur earlier. 
There are several reasons for which an earlier breakup of the severed freely expanding rim would be observed. 
First\mydeleted{ and foremost}, the rim already inherited well-formed corrugations from the sheet pre severance (see \S\,\ref{IV.3} also), with significant local fluctuations in the thickness relative to the average rim thickness, favoring local pinch-off.  
\myadded{Thus, a comparison to theory is most meaningful when considering the time instance for full fragmentation instead of the first instance of fragmentation. }
Furthermore, given that the rim is severed with constant volume $v_{\textrm{r}}\sim r_{\textrm{s}} b^2$, an increase in $r_{\textrm{s}}(t)$ of approximately 50\% [e.g., see fig.\,\ref{fig:3}(e)], for example, leads to a corresponding decrease in $b$, and with it, a decrease in fragmentation time $t_{\textrm{br}}$ up to $\approx26\%$. 
In sum, in the absence of liquid flow from the sheet, the thickness of the rim cannot self-adjust to fulfill the Bo $=1$ criterion and the rim destabilizes into fragments \emph{mainly} subjected to capillary breakup via RP instability with initial strong fluctuations in thickness inherited from, and an enhanced thinning due to, the inherited ballistic radial expansion.

\subsection{Number of elements: Ligaments and fragments}
\label{IV.3}

In figs.\,\ref{fig:5}(a-e) we present the population of ligaments, $N_{\textrm{l}}$ and fragments, $N_{\textrm{f}}$, resulting from the breakup of the rim for two different droplets, $d_0$=30 and $40\,\mu$m and several $\textrm{We}_\textrm{d}$. 
We carefully select the frames that allow for clear identification of fragments that come from the destabilization of the rim, without counting those that come from the breakup of the ligament tips or those ejected from the inner sheet at a later time (see \S \ref{IV.4}). 
Note that the data points that appear clustered correspond to different moments of detachment. 
It is important to recall that ligaments and fragments are quantified at different moments, even though the physics is set by the time of detachment at VP-impact. In other words, in most cases we count ligaments from the first frame (recall the double-framing camera setup) as they are formed before rim severance, while the fragments are revealed later, seen predominantly in the second frame. 
Therefore, the points that represent the number of ligaments and fragments from the same vaporization times are shifted in time.  
Clearly, the resulting fragment population is larger than that of the ligaments, with a near-constant ratio of 2 < $N_{\textrm{f}}/N_{\textrm{l}}$ < 3 with a slight increase of this ratio for higher $\textrm{We}_\textrm{d}$. 
This dependence on $\textrm{We}_\textrm{d}$ appears to be weaker in the case of ligaments. 
In order to further describe our observed population of elements, we need to first discuss their origin. 
This is done next. 

\subsubsection*{Ligaments} \label{sec:Ligaments}
The corrugation of the rim is the consequence of a nonlinear interplay between RT and RP instabilities, where the preferential wavelength (corresponding to the fastest-growing mode) is close to the one defined by the dispersion relation of the RP instability \citep{wang_bourouiba_2018_rim}. 
Some protuberances grow into elongated ligaments and others do not. 
The physical constraint that determines the growth of corrugations into ligaments is a combination of local inertial forces induced by the rim deceleration, the capillary-driven retraction of the sheet, and the flow of the liquid into the rim from the sheet  \myreplaced{\mbox{\citep{klein_drop_2020,wang_bourouiba_2021_growth}}}{\mbox{\citep{wang_bourouiba_2021_growth}}}. These authors established that such influx of liquid from the sheet is not sufficient to support ligament growth for all emerging corrugations and derived a criterion of ligament growth from some corrugations, while others remain frozen. 
The authors proposed an analytical solution to quantify the population of the ligaments on a corrugated rim as a function of time and the impact $\textrm{We}$.
Furthermore, \citet{wang_bourouiba_2021_growth} quantifies the resulting population of ligaments as the ratio of the unsteady evolution of the rim perimeter and the minimum distance between two adjacent ligaments, $\lambda_m$. 
The value of $\lambda_m$ evolves over time and is a function of the sheet thickness' radial distribution as well as the balance between the incoming rate of liquid fed into the rim from the sheet and the rate of liquid volume shed from the ligaments  (i.e., the flow rate from the ligaments). 
The analytical definition of the number of ligaments from \citet{wang_bourouiba_2021_growth} reads

\begin{equation}
    N_{\textrm{l}}=\frac{2\pi R_{\textrm{s}}(T)}{\Lambda_{\textrm{m}}(T)}=\frac{8\sqrt{3}Q_{\textrm{out}}(T)}{W^{3/2}(T)\beta(T)}=\Theta(T)(\textrm{We}_\textrm{d}/4)^{3/8},\,\,\, \beta(T)=\sqrt{\alpha(T)^2+\frac{5}{2}\alpha(T)-2}.
\label{Eq:3}
\end{equation}

Here, $R_{\textrm{s}}(T)\equiv r_{\textrm{s}}(T)/d_0$, $\Lambda_m = \lambda_{\textrm{m}}/d_0(T)$ is the non-dimensional minimal distance between ligaments, $Q_{\textrm{out}}(T)$ is the flow rate from the ligaments, $W(T)$  is the non-dimensional average ligament width, and $\alpha(T)$ is the ratio between the rim thickness and ligament width. 
These functions are further detailed in Appendix\,\ref{A.3}. 

\begin{figure}
\centering
\includegraphics[width=1\linewidth]{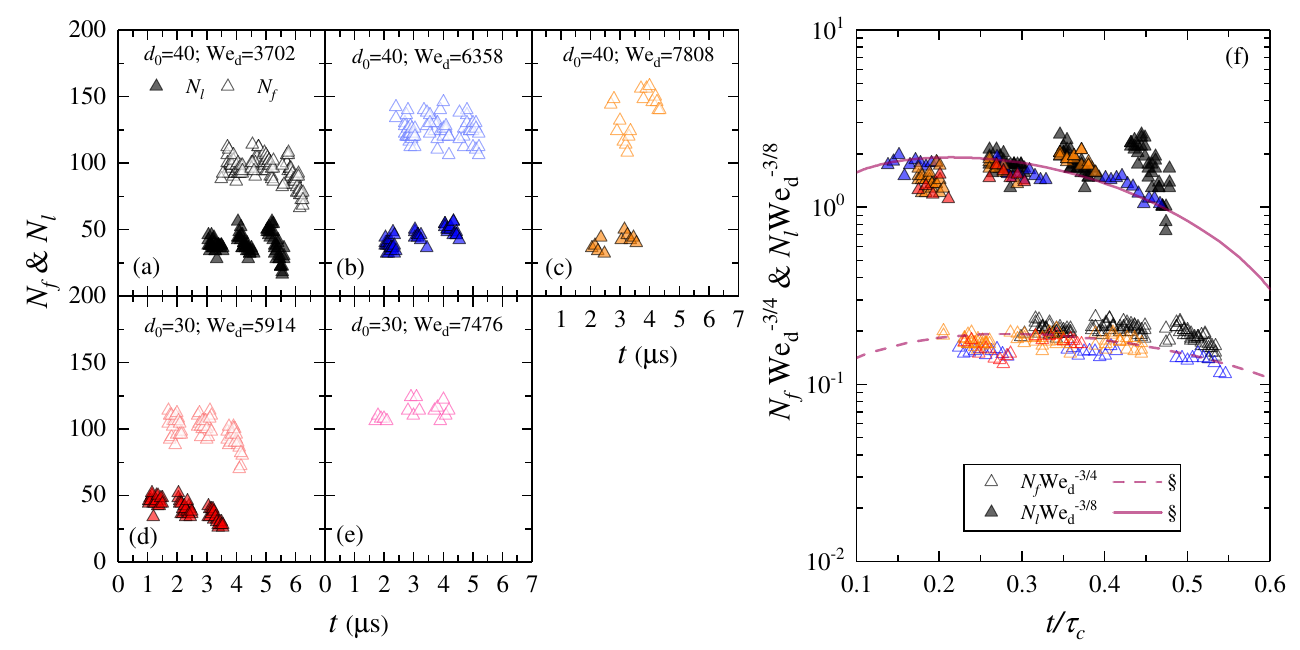}
\caption{Number of elements post rim severance. 
(a-e) Number of ligaments (closed points) and fragments (open points) for different $d_0$ and $\textrm{We}_\textrm{d}$ at several detachment times. 
For the $d_0=30\,\mu$m and $\textrm{We}_\textrm{d}$=7476 case, the number of ligaments cannot be quantified since the breaking time is too short. 
(f) Scaled number of elements with their corresponding $\textrm{We}_\textrm{d}^{\alpha}$ scaling, where $\alpha=-3/8$ for ligaments and $\alpha=-3/4$ for fragments, as a function of non-dimensional time $t/\tau_{\textrm{c}}$. 
The predictions \myadded{of \ref{Eq:3}) and (\ref{Eq:4})}\myadded{, with no fitting being performed,} for ligaments and fragments are shown with solid and dashed lines.\mydeleted{, corresponding to (\ref{Eq:3}) and (\ref{Eq:4}), respectively.} 
}
\label{fig:5}
\end{figure}

Recall that we make use of $\textrm{We}_\textrm{d}$ (see the discussion in  \S \ref{sec:exp}).  
In fig. \ref{fig:5}, we quantify the number of ligaments from different instances of rim severance. 
At low deformation $\textrm{We}_\textrm{d}$ we observe a slight decrease in the number of ligaments after rim detachment [see fig. \ref{fig:5}(a,d)]. 
In the absence of liquid input, \myadded{we hypothesize} \mydeleted{it is expected} that \mydeleted{the} small ligaments retract driven by capillary tension from the remaining free\mydeleted{-flying} liquid of the rim \mydeleted{[see fig. \ref{fig:2}(a)]}. 
\myadded{We note that it is a hypothesis that in the absence of liquid input, small ligaments retract driven by capillary tension from the remaining free liquid of the rim. We note that here, the relatively large 650\,ns time difference between the two frames does not allow for direct unambiguous identification of such retraction. Furthermore, longer ligaments (especially at late expansion time) break up on this timescale and the counted ligament number may decrease as a result. These two effects combined could lead to an effective decrease in the population over time post-rim severance.  }
\myadded{These two effects may contribute} \mydeleted{This retraction leads} to an effective decrease in the population over time after rim severance, as can be clearly appreciated in figs. \ref{fig:5}(a,d).
 
In fig. \ref{fig:5}(f) we illustrate the number of ligaments rescaled by $\textrm{We}_\textrm{d}^{3/8}$, and compare the experimental data with (\ref{Eq:3}), which is computed (see Appendix \ref{A.3}) based \myadded{only} on the \mydeleted{fitted} sheet expansion coefficients for tin.  
\myadded{No fitting of the equations to the ligament or ligament-fragment data is performed. }
The resulting predictions for the number of ligaments are depicted in fig.\,\ref{fig:5}(f). 
We observe  agreement between our experimental data and the model predictions with the thickness profile from the current work. 

\subsubsection*{Fragments}
Regarding the number of fragments, we observe their origin to be inherited from corrugation of the rim prior to severance from the unsteady sheet. 
Similarly to the observed decrease in the number of ligaments (especially at low deformation Weber number) as shown in fig. \ref{fig:5}(a), the population of fragments resulting from the complete breakup of the rim after severance decreases over time. 
As the liquid influx from the sheet ceases, at low $\textrm{We}_\textrm{d}$, \myadded{in some cases, the rim undergoes a rapid capillary reconfiguration due to the bulge scavenging}. This effect is accentuated at late time in the mentioned cases where the capillary-driven retraction of the sheet provides a better condition for bulge merging [see fig. \ref{fig:5}(a,d)]\myadded{: at late times $t/\tau_{\textrm{c}}$, the expansion velocity of the rim is small and breakup times large}. 
At a similar timescale, the rim undergoes an irreversible capillary-driven fragmentation which we quantify in fig. \ref{fig:4} to be on the order of  $\tau_b\approx500\,$ns, which is compatible with a corresponding capillary timescale $\tau_b=(\rho b^3/6\sigma)^{1/2}\approx 300\,$ns for a rim of $b\approx5\,\mu$m as noted above. 
We next postulate that the \emph{average} population of fragments is determined by the inherited corrugation number $N_{\textrm{c}}$. 
Thus, in the case of the number of fragments, we can consider that the \emph{average} population of fragments $N_{\textrm{f}}=N_{\textrm{c}}$. 
A complete analytical solution for $N_{\textrm{c}}$ was derived by \citet{wang_bourouiba_2021_growth}, showing that the average distance between two bulges is indeed captured well by the wavelength that would emerge from the fastest growing mode of the RT-RP instability with the Bo $=1$ condition, with $N_{\textrm{c}}=4\pi R_{\textrm{s}}(T)/9B(T)$, where, again, $R_{\textrm{s}}(T)\equiv r_{\textrm{s}}(T)/d_0$ and $B(T)\equiv b(T)/d_0$ are the non-dimensionalized sheet radius and rim thickness, respectively. 
Combining this expression with the temporal evolution of the sheet, adjusted to the laser-impulse sheet [see (\ref{Eq:1})], we deduce the description for the time evolution of the corrugation wavenumber in the system to read 

\begin{equation}
    N_{\textrm{c}}=\frac{4\pi}{9}\left(\frac{\textrm{We}_\textrm{d}}{4}\right)^{3/4} \cdot\frac{(b_3(T-T_{\textrm{m}})^3+b_2(T-T_{\textrm{m}})^2+b_0)}{[-6(3b_3(T-T_{\textrm{m}})+b_2)]^{-1/2}}.
\label{Eq:4}
\end{equation}

Taking into account the postulated equivalence $N_{\textrm{c}}=N_{\textrm{f}}$, we show the scaled $N_{\textrm{f}}$ with its corresponding dependence on $\textrm{We}_\textrm{d}$ in fig. \ref{fig:5}(f). 
Good accordance is found between the experimentally measured fragment population and the predictions based on (\ref{Eq:4}). 
\myadded{Again, no fit is performed to the fragment data. }
\myreplaced{This agreement between data and model }{Closer observation} suggests \myadded{that, overall,} no significant scavenging \citep{wang_bourouiba_2021_growth} of corrugations occurs in the short time window between rim detachment and full fragmentation.

\subsection{Inner sheet behavior}\label{res:dynamics:innersheet}
\label{IV.4}

\begin{figure}
\centering
\includegraphics[width=1\linewidth]{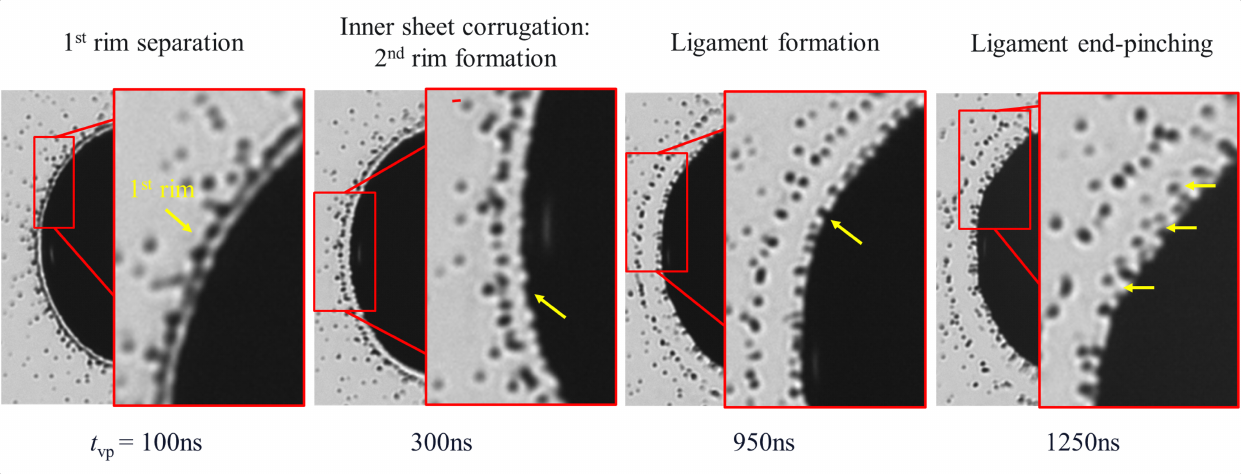}
\caption{Inner sheet behavior after rim release. 
The images are obtained with a droplet of $d_0=30\,\mu$m and the pre-pulse laser energy $E_{\textrm{pp}}$=22\,mJ after $2\,\mu$s upon laser-droplet interaction. 
From left to right, the shadowgraphs disclose the time evolution of the inner sheet: rim detachment and separation from the remaining sheet ($t$=100\,ns, \myadded{note that the vaporization of the neck and, thus, the actual release of the rim already occurs on the 5-ns time scale of the VP}), initial corrugation of the second rim ($t$=300\,ns), ligament formation ($t$=950\,ns), and their further fragmentation via end-pinching events ($t$=1250\,ns).}
\label{fig:6}
\end{figure}

After the rim detachment, the expanding sheet develops a second bounding rim. 
The periphery of the sheet remains stable for a certain interval of time, and later the second rim manifests itself as a growing set of small corrugations. Afterward, some corrugated bulges eventually lead to the formation of elongated ligaments. 
Finally, these ligaments destabilize and break into droplets through end-pinching (see fig. \ref{fig:6}). 
Based on our arguments from \S \ref{IV.1} on the rim and sheet expansion, we expect that the second rim will remain stable for a certain period of time before reaching a critical thickness at which ligament formation is possible. 
At that point, the liquid inflow from the sheet into the rim, governed by the unsteady non-Galilean local Taylor-Culick law \citep{wang2023non}, enables the cycle to re-start: the thickness of the rim adjusts to remain close to the local and instantaneous capillary length, meeting the Bo $=1$ criterion.  As long as this criterion holds, the corrugation is governed by nonlinear RT-\myadded{R}P instabilities, with the corresponding wavenumber.

\subsubsection*{Rim re-formation  timescale}

We may estimate the timescale for the rim \emph{re-formation} by acknowledging that the clearest visible manifestation of local capillary equilibrium of the rim diameter that fulfills the Bo $=1$ criterion is the onset of the formation of ligaments \citep{wang_bourouiba_2018_rim}. 
This realization enables us to determine the local diameter $b$ of this new rim through $\textrm{Bo}=1 \rightarrow b(t) = [\sigma/(-\ddot{r}_{\textrm{s}}(t)\rho)]^{1/2}$; the volume contained in this cylinder is given by $(2 \pi r_{\textrm{s}}(t))(\pi b^2/4)$ and taking the incoming liquid speed to be  $v(t) \sim [2\sigma/\rho h(t)]^{1/2}$\myreplaced{, the local Taylor-Culick velocity \mbox{\citep{Culick1960}} as applied to the frame of the rim following \mbox{\citet{wang2023non}},}{\mbox{\citep{wang2023non}}} where $h(t)$ is the local thickness of the sheet.
 
Thus, it would take a time interval on the order of $\Delta t = \pi b^2/(4 v h)$  for the liquid to flow into the rim through an area spanning $2 \pi r_{\textrm{s}}(t) h(t)$, assuming no significant change in $b(t), v(t)$, or $h(t)$ over the filling of the new rim, i.e. if $\Delta t\ll \tau_{\textrm{c}}$. 
We note that this refilling timescale is independent of  $\textrm{We}_\textrm{d}$ if $\textrm{We}_\textrm{d}\,\gg 1$, and is a function only of dimensionless time $t/\tau_{\textrm{c}}$. 
In addition, we account for an offset inertial timescale $t_{\textrm{i}}$ required to establish the local sheet velocity influx from \citet{wang2023non}, as approximated by $t_{\textrm{i}} \approx (\rho \Omega_\textrm{tip} / \sigma \pi)^{1/2} = (\pi r_{\textrm{s}} h^2 \rho /2\sigma)^{1/2}$, where $\Omega_\textrm{tip}$ is the volume of the tip prior to rim formation, considered as a time-varying cylinder with diameter equal to local thickness $h$ \citep{Keller1983_TC,Taylor-Culick-approximation_2020}. 
We may expect that the rim re-formation takes place on a timescale of the order of $t_{\textrm{r}}\sim \Delta t + t_{\textrm{i}}$ independent of We for the current case where $\textrm{We}_\textrm{d}\gg1$. 

\begin{figure}
\centering
\includegraphics[width=1\linewidth]{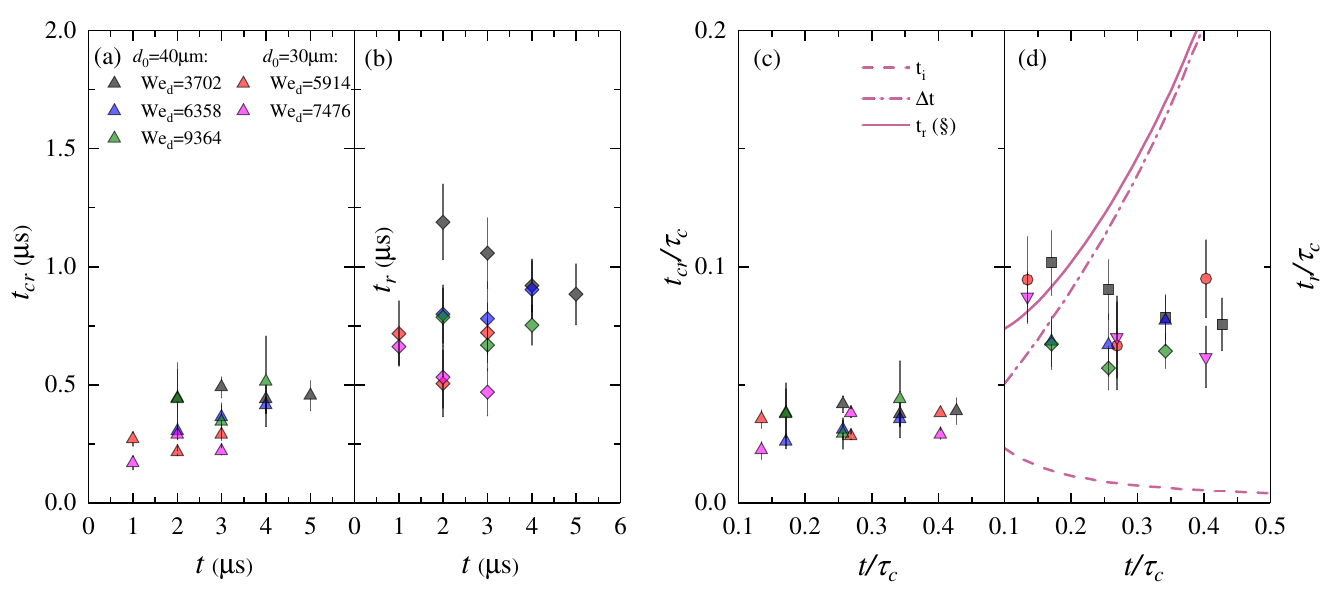}
\caption{Second rim re-formation post first rim severance. 
(a) Inner sheet corrugation time $t_\textrm{cr}$ for various $d_0$ and $\textrm{We}_\textrm{d}$. 
(b) The onset times for the formation of inner sheet ligaments $t_r$ (equaling the timescale for rim re-formation, see the main text) for the same data set as in (a). 
(c) Corrugation time of the inner sheet scaled with capillary time, $t_\textrm{cr}/\tau_{\textrm{c}}$ over non-dimensional time $t/\tau_{\textrm{c}}$. 
(d) Ligament formation time of the inner sheet scaled with capillary time, $t_{\textrm{r}}/\tau_{\textrm{c}}$ over non-dimensional time $t/\tau_{\textrm{c}}$. 
The dashed line corresponds to the inertial time $t_{\textrm{i}}$, the dash-dotted line is the time required to refill the rim $\Delta{t}$, and the solid line is the sum of both, rim re-formation  time $t_r=\Delta{t}+t_{\textrm{i}}$. \myadded{No fitting is performed.}  
} 
\label{fig:7}
\end{figure}

Figure \ref{fig:7}(a) shows the timescale of the formation of corrugations on the inner sheet edge, following rim release, and fig. \ref{fig:7}(b) the timescale for the formation of ligaments on the new rim. 
Times were obtained by identifying the first frame in which significant corrugations and ligaments were visible and the last frame in which corrugations and ligaments were observed along the entire half of the sheet (following the same process as in \S \ref{IV.2} and \ref{IV.3}) and taking the average of these times as input. 
We note that the experimental results here are limited by the finite resolution of the optical system: corrugations much smaller than the resolution limit are not visible. 
Thus, our smaller scale of observation is an upper limit at the relevant timescale. 
This underlying systematic offset becomes more significant at higher We as corrugations are ever smaller. 
This technical limitation still holds for the identification of ligaments, but it is less acute given their larger size. 
Experimental challenges notwithstanding, in figs. \ref{fig:7}(c) and \ref{fig:7}(d) we show the scaled corrugation and re-formation times $t_\textrm{cr}/\tau_{\textrm{c}}$ and $t_{\textrm{r}}/\tau_{\textrm{c}}$ as a function of non-dimensional time, $t/\tau_{\textrm{c}}$; the scaling leads \mydeleted{to the expected} \myadded{a modestly} reduced spread in the data. 
\mydeleted{These experimental data may be contrasted with the model predictions.} 
The model gives a comparable order of magnitude estimate for the rim re-formation timescale and no dependence on We, and with a better capture of the experimental values \myreplaced{only at early time}{at early time only however}. 
\myadded{
We hypothesize that the faster-than-expected timescale to shedding of the new rim at late severance times could have its origins in the inherited corrugations. The inherited corrugations point to (i) a potential higher liquid influx into the new rim at the locations of the original corrugations, due to a suction-inducing inherited local curvature, and (ii) a  triggering of instabilities  that may lead to shedding earlier than the typical time needed for the thickness  to fulfill the \mbox{Bo $=1$} criterium. The effects of such inherited corrugations will increase with time given the increase in rim and rim-corrugation diameter with time. }

\section{Application perspective}\label{sec:appl}
EUV light as used in nanolithographic applications is currently produced by the laser-irradiation of so-called "mass-limited" liquid tin microdroplets \citep{mizoguchi2015performance,fomenkov2017light,versolato2019physics}.
In this application, thin liquid tin sheets are produced from droplets using a \SI{}{\nano\second} laser pre-pulse \citep{Klein2015,Hudgins2016, Gelderblom2016,kurilovich2016plasma, kurilovich2018power, liu_2020_mass, liu_2023_mass} and serve as targets for an energetic main pulse (MP) to efficiently produce EUV-emitting plasma \citep{fomenkov2017light, versolato2019physics}.
%
For industrial use, it is important to minimize microscopic liquid debris that may coat nearby light collection optics and impact machine uptime. 
\citet{liu_2023_mass} suggested minimizing the expansion time required to reach a set target size, which, in turn, would minimize the mass lost from the sheet via ligaments and ligament-fragments at the time of the MP impact.  
However, even in such optimized prepulse scenarios, the rim accumulates significant volume, impacting the homogeneity of the plasma generation with a massive $\approx 1\mu$m rim bounding a 10--100 nm thin sheet.
This study now suggests a path to minimizing the production of ligament-fragments outside the caustic of the main pulse and improving the homogeneity of the target volume. 
We propose exploiting the rim severance and re-formation discussed herein, with a sequence of vaporization pulses, with rim severance at each pulse.
The first such \mydeleted{a} pulse would severe the first rim just prior to ligament-fragment shedding, thus avoiding liquid debris outside of the MP caustic.
The spatial profile of the laser pulse could be optimized to vaporize only the liquid connecting with the rim, e.g., using a doughnut shape.
As our research shows, rapid destabilization of the disconnected rim does not lead to  significant excess radial redistribution (i.e., outside of the MP caustic) of the mass contained therein.
We further demonstrated that a new rim rapidly re-forms upon  detachment of the previous one, and any new ligament-fragments formed will occupy the space between old and new rim -- and thus remain in the main pulse laser caustic and not add to debris generation.
To still improve the homogeneity of the target mass delivery, for example for the second, or third cycles, vaporization pulses would disconnect the recurring rims. This approach would enable to optimally distribute the tin for generating EUV light. Note that the vaporization pulses could be spaced by time $t_r$ intervals matching the rim re-formation as well.

\section{Conclusions}
Laser interaction with tin microdroplets leads to an unsteady sheet that radially expands,  bounded by a rim of time-varying thickness. 
The dynamics of the sheet, as well as the formation of the ligaments and their subsequent fragmentation, are governed by the unsteadiness of the rim destabilization, itself highly dependent on the  liquid inflow from the sheet into the rim. 
Therefore, severing the rim from the expanding sheet, proposed here, is a novel method to comprehensively study the resulting hydrodynamics in the absence of the mentioned liquid inflow. 
Here, we aimed to answer the following questions: 
 \begin{enumerate}
 \item Can we create the conditions of a severed rim to further clarify the role of fluid influx into a pre-formed rim?  In particular, we aim to determine the effect of interruption of such influx on the evolution of a freely evolving toroidal interfacial structure in radial expansion. 
\item  If so, how does the number of rim corrugations and ligaments evolve upon severance of the rim from the sheet, i.e., upon interruption of fluid influx from the sheet into the rim?
\item What is the final fragment number, or associated wavenumber,  from such a severed isolated rim in expansion? In particular, does it continue to follow the scaling laws inherited prior to severance or do spontaneous capillary re-configurations of the corrugations change the law, i.e., final fragment number?
\item On which timescale would a new rim re-from on a radially expanding sheet and how does it depend on the stage of the unsteady sheet expansion?
 \end{enumerate}

First, we indeed established a system where we could apply a severance of the rim from the sheet at a desired time of the sheet expansion, and then follow the subsequent evolution of the rim, post severance. 
This was done by laser-vaporizing the connecting thin neck that bridges the rim with the sheet\myadded{, with negligible influence of the laser pulse on the much thicker rim itself}. 
Second, upon detachment, we observed a ballistic radial expansion of the free-flying rim subjected to a fast capillary-driven breakup. 
We find that in addition to the thinning of the rim induced by its radial expansion absent of inflow from the sheet, the volume is also further drained by the shedding of fragments from the inherited ligaments. 
We find that the rim breakup occurs mostly at the regions of reduced thickness that are inherited from the corrugated rim pre severance. 
Indeed, the unbalance in liquid mass inflow into the rim and radial expansion enhance the capillary pinch in the relatively thinner regions of the rim, with minimal reconfiguration or scavenging occurring. 
Given that the local curvatures at the foot of the inherited ligaments and the non-ligament corrugations are comparable, the timescales of pinch-off are also comparable, leading to a simultaneous fragmentation of the inherited ligament bases and inter-ligament corrugations. 
The later we severed the rim, the thicker it is, and the longer it takes post severance to fragment, with the correction that the rim post severance thins given its ballistic radial expansion, and is moreover already corrugated, two factors contributing to earlier fragmentation than a smooth, non-elongating rim.  
Third, we quantitatively analyzed the population of ligaments and the resulting fragments post-rim full fragmentation. 
We show that the inherited number of corrugations, i.e. their wavenumber, inherited from the sheet prior to rim severance corresponds to the number of fragments, i.e. $N_{\textrm{c}}=N_{\textrm{f}}$. 
Furthermore, we note that the absence of liquid influx, coupled with its thinning due to radial expansion, appears to prevent any further scavenging or merging of corrugations or ligaments. 
As a result, we also recover a number of ligaments and non-ligament fragments that are compatible with those predicted in the context of unsteady dynamics of droplet impact on pillars of comparable size, with  $N_{\textrm{l}}\sim \textrm{We}_\textrm{d}^{3/8}$ and $N_{\textrm{f}}\sim \textrm{We}_\textrm{d}^{3/4}$ when using as inputs the sheet's expansion trajectory and a consistent sheet thickness profile \myadded{that is derived from it}. 
The equivalence $N_{\textrm{c}}=N_{\textrm{f}}$ enables us to quantify the corrugation wavenumber in laser-driven tin sheets from rim detachment, given the experimental limitations, this wavenumber cannot be easily quantified otherwise. Note that we find further motivation to experimentally revisit the thickness profile most relevant to the laser-tin case.

Finally, we explored inner sheet's late-time dynamics and highlighted several features. 
We correlated the onset of the ligament formation with the moment of rim re-formation, requiring some time on the order of the local inertial timescale.  
Subsequently, the inner sheet follows an expansion trajectory close to the original one, with re-formation of a new rim that is subsequently prone to corrugation and ligament formation. 
To account for this unsteadiness, we proposed  a first-order model that captures the order of magnitude of the  onset time of the inner sheet rim re-formation at early sheet evolution time. 
In summary, our study sheds light on the dynamics of rim fragmentation in absence of liquid inflow from its original sheet; establishes a method to quantify the corrugation wavenumber in laser-driven tin sheets; and examines the re-formation of the rim bounding the sheet. 
Finally, the rim re-formation redistributes mass on the sheet, allowing interesting novel strategies for an optimized target design to potentially further improve industrial nanolithography. 

\begin{acknowledgments}
This work was conducted at the Advanced Research Center for Nanolithography (ARCNL), a public-private partnership between the University of Amsterdam (UvA), Vrije Universiteit Amsterdam (VU), Rijksuniversiteit Groningen (UG), the Dutch Research Council (NWO), and the semiconductor equipment manufacturer ASML and was partly financed by ‘Toeslag voor Topconsortia voor Kennis en Innovatie (TKI)’ from the Dutch Ministry of Economic Affairs and Climate Policy. The authors were supported, in part, by funding from the European Research Council (ERC) under the European Union’s Horizon 2020 research and innovation programme under grant agreement No 802648. LB acknowledges support from the USDA, Inditex, the National Institutes of Health, Analog Devices, and the National Science Foundation.
\end{acknowledgments}

\section*{Data Availability Statement}
The data that support the findings of this study are available from the corresponding author upon reasonable request.

\section*{Competing interests}

The authors are not aware of any conflict of interest that might affect the objectivity of this study.

\section*{Author ORCID}

\noindent M. Kharbedia, \url{https://orcid.org/0000-0002-2128-9945}

\noindent B. Liu, \url{https://orcid.org/0000-0001-7122-6283}

\noindent R. A. Meijer, \url{https://orcid.org/0000-0003-1738-4488} 

\noindent D. J. Engels, \url{https://orcid.org/0000-0001-7363-8716}

\noindent H. K. Schubert, \url{https://orcid.org/0009-0000-4499-0091}

\noindent L. Bourouiba, \url{https://orcid.org/0000-0001-6025-457X}

\noindent O. O. Versolato, \url{https://orcid.org/0000-0003-3852-5227}

\newpage

\appendix

\mydeletedappendix{Ballistic trajectory for the severed rim}\label{A.2}


\mydeletedred{
We find a ballistic trajectory for the severed rim with a constant speed close to the speed of the rim upon severance from the sheet. 
Figure\,\ref{fig:A2} provides a more detailed view of the velocity of the severed rim in comparison with the connected rim that can be obtained from $r_{\textrm{s}}(t)$ from the third-order polynomial fit used to determine the sheet expansion (see \S \ref{IV.1}). 
No significant acceleration of the detached rim is apparent. 
Suboptimal experimental conditions, with limited data available on the time axis and a possibly higher than average VP pulse energy for the $d_0$=40\,$\mu$m, We$_d$=3702 case in panel (a) led to large scatter in the data with corresponding large uncertainties. 
We also note that the sheet radius evolution description (Fig.\,\ref{fig:3}) has increasing error as we approach the  early-time stage of sheet expansion, due to initial conditions of sheet formation not being fully captured by the model, as discussed elsewhere. Hence, it is possible that a future correction incorporating early sheet expansion would be compatible with a faster early time expansion than current encompassed by our model.}

\section{Coefficients for sheet expansion and number of ligaments}
\label{A.3}

\subsection{Coefficients of sheet expansion}

To analytically derive the equation for the temporal evolution of the rim, we recall the theoretical treatment based on \citet{wang2023non} that takes as input the thickness profile. In the intermediate-Weber regime of $250 < \textrm{We}< 10000$, the authors detail the combined effect of the continuous thinning of the sheet and fragment shedding from the ligaments to establish a non-Galilean Taylor-Culick law that accounts for the rim expansion,  reading 

\begin{equation}
    -6H(R_{\textrm{s}},T)\left(\frac{R_{\textrm{s}}(T)}{T}-\dot{R}_{\textrm{s}} (T)\right)^2+\left(2-\frac{\pi}{7} \right)=0,
\label{Eq:C1}
\end{equation}
where $H(r,T),\,R_{\textrm{s}}(T),\,\dot{R}_{\textrm{s}}(T)$ are the non-dimensionalized sheet thickness, sheet radius and expansion velocity, respectively, as functions of non-dimensional time $T=t/\tau_{\textrm{c}}$. 
The theoretical derivation using eqs.\,(A1-A18) in \citet{wang2023non} arrives at the dimensionless thickness equation in the case of the water-pole impact: 
\begin{equation}
    \begin{aligned}
        H_{\textrm{wp}}(R_{\textrm{s}},T) &= \frac{T\sqrt{6\textrm{We}}}{6a_3R_{\textrm{s}}^3+a_2R_{\textrm{s}}^2T(6\textrm{We})^{1/2}+a_1R_{\textrm{s}}T^2\textrm{We}}, \\
            \end{aligned}
\label{Eq:C2}
\end{equation}
\noindent with coefficients labeled $\left\{a_1,a_2,a_3 \right\}_\textrm{wp}$ with $\left\{a_1=24.4(2), a_2=-38.1(4), a_3=35.2(3)\right\}_{\textrm{wp}}$,  provided in \citet{wang_bourouiba_2017_thickness}. 
An alternative functional form for the thickness profile was heuristically introduced by \citet{liu_2020_mass} for the laser-tin droplet impact case, introducing an additional $a_0$ coefficient while dropping the $a_3$ term to better describe the observed profile. 
This alternative form reads
\begin{equation}
    \begin{aligned}
        H_{\textrm{ef}}(R_{\textrm{s}},T) &= \frac{T\sqrt{6\textrm{We}}}{a_2R_{\textrm{s}}^2T(6\textrm{We})^{1/2}+a_1R_{\textrm{s}}T^2\textrm{We}+a_0T^3 6^{-1/2}\textrm{We}^{3/2} },
    \end{aligned}
\label{Eq:C3}
\end{equation}
with coefficients $\left\{a_0=1.65(2), a_1=6.9(3), a_2=-2.4(8)\right\}_{\textrm{ef}}$ \citep{Liu_Bo_phdthesis, liu_2020_mass} empirically determined from a direct fit to experimental data. 
We recall the derivations by \citet{wang2023non} of the $\left\{b_0, b_2, b_3\right\}$ coefficients in the governing equation of the sheet expansion (\ref{Eq:1}) with the following change in variables: 

\begin{equation}
    Y=R_{\textrm{s}} \sqrt{\frac{6}{\textrm{We}}},\,\,U=T-T_{\textrm{m}}.
\label{Eq:C4}
\end{equation}

\noindent  Substitution of (\ref{Eq:C2}) or (\ref{Eq:C3}) with (\ref{Eq:C4})  into (\ref{Eq:C1}), gives

\begin{equation} \label{Eq:C5}
    \begin{split}
        6\left(Y-\dot{Y}\left(U+T_{\textrm{m}}\right)\right)^2 &= \left(2-\frac{\pi}{7}\right) \Big[ a_3Y^3\left( U+T_{\textrm{m}}\right)  + a_2Y^2\left( U+T_{\textrm{m}}\right)^2 \\
        &\quad + a_1Y\left(U+T_{\textrm{m}}\right)^3 + a_0\left( U+T_{\textrm{m}}\right)^4 \Big].
    \end{split}
\end{equation}

\noindent This complete form includes all four $a_{\textrm{i}}$ parameters. With the newly defined variables, (\ref{Eq:1}) from the main text can be re-expressed as 
\begin{equation}
    Y(U)=\beta_0+\beta_2U^2+\beta_3U^3,
\label{Eq:C6}
\end{equation}
and the coefficients $b_0,\,b_2,\,b_3$ are related to $\beta_0,\,\beta_2,\,\beta_3$, which are explicit functions of $a_{\textrm{i}}$, via

\begin{equation}
      \left( \beta_0,\,\beta_2,\,\beta_3 \right)=\sqrt{6}\left(b_0,\,b_2,\,b_3 \right).
\label{Eq:C7}
\end{equation}

\begin{figure}
    \centering
    \includegraphics[width=0.75\linewidth]{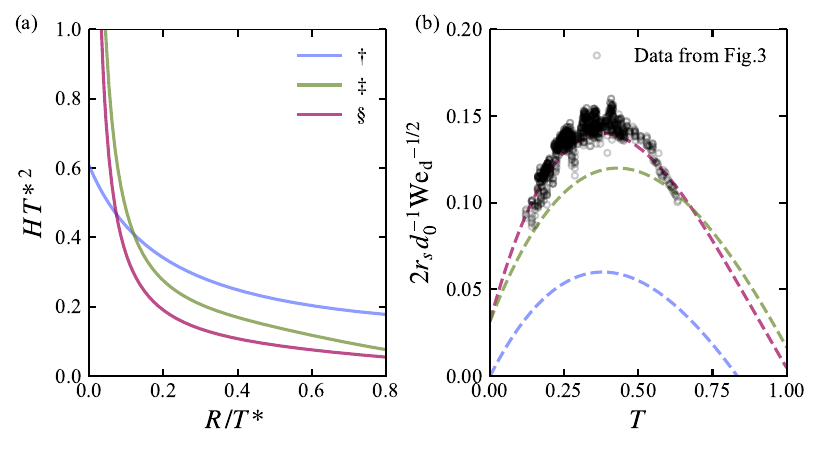}
    \caption{Thickness profile and sheet expansion for three consistent sets of coefficients $\left\{a_{\textrm{i}} \right\}$ and $\left\{b_{\textrm{i}} \right\}$. (a) Thickness profiles with $\left\{a_0=1.65(2), a_1=6.9(3), a_2=-2.4(8)\right\}_{\textrm{ef}}$ ($\dag$),  $\left\{a_1=24.4(2), a_2=-38.1(4), a_3=35.2(3)\right\}_{\textrm{wp}}$ ($\ddag$) and $\left\{a_1=28.6(2), a_2=-26.2(1), a_3=23.5(3)\right\}_{\textrm{tw}}$ ($\S$), from empirical fit, water on pole and the profile derived in this work, respectively (see the main text). 
    Both axis are expressed in similarity variables, with $H$, $T^*=\dot{r}_0t/2d_0$, and $R$ being dimensionless thickness, time, and radial coordinate, respectively. 
    (b) Sheet expansion trajectories obtained from the three sets $\left\{b_0,\,b_2,\,b_3\right\}$ consistent with the thickness profiles in (a) along with experimental data set as seen in fig.\,\ref{fig:3} for $d_0=30, 40\,\mu\textrm{m}$. } 
    \label{App:Thickness-Expansion}
\end{figure}

\noindent After substituting (\ref{Eq:C6}) into (\ref{Eq:C5}), algebraic manipulation enables us to derive $\left\{b_{\textrm{i}} \right\}$ in terms of $\left\{a_{\textrm{i}} \right\}$, and vice versa.

In the following, we use as input two thickness profiles: (i) from water pole impact by  \citet{wang2023non} (labeled "wp") and (ii) from the empirical fit from \citet{liu_2020_mass} (labeled "ef"), see fig.\,\ref{App:Thickness-Expansion}(a). 
These two profiles lead to two sets of corresponding expansion curve coefficients as input to solve (\ref{Eq:C1}). 
For (iii) a third case (labeled "tw") we reverse the approach and obtain thickness profile parameters from experimentally determined sheet expansion coefficients. 
Note that in the current study of laser impact on tin, the experiments give $T_{\textrm{m}}=0.38$, contrary to $T_{\textrm{m}}=0.43$ for the water pole impact. 
Next, we note that the consistent substitution of $\textrm{We}=\textrm{We}_\textrm{d}/4$  (see the main text) in (\ref{Eq:C2}-\ref{Eq:C4}) does not modify (\ref{Eq:C5}). 
Also, we note that the identity $T=\sqrt{\textrm{We}/6} T^*$ from equation (2.3) in  \citet{wang2023non} is maintained in the main text with the aforementioned substitution of We number, along with the definition of the inertial time $T^*$ in terms of $\dot{R}_{\textrm{s}}/2$ instead or $U$.  

The three sheet expansion curves are shown in fig.\,\ref{App:Thickness-Expansion}(b). 
The first curve is based on the water on pole thickness profile ($\ddag$), with expansion coefficients $\left\{b_0=0.12, \,b_2=-0.41, \,b_3=0.16\right\}_{\textrm{wp}}$. 
The second is derived from thickness profile from the empirical fit ($\dag$) from \citet{liu_2020_mass}, resulting in derived coefficients $\left\{b_0=0.06, \,b_2=-0.36, \,b_3=0.15\right\}_{\textrm{ef}}$. 
Finally, the third curve represents the expansion used in this work ($\S$) that fits our experimental data, with its corresponding coefficients $\left\{b_0=0.14, \,b_2=-0.58, \,b_3=0.43\right\}_{\textrm{tw}}$. 
The corresponding thickness profile coefficients were reported in the main text and are not repeated here.  
For comparison, we include experimental data of sheet expansion for the $d_0=30, 40\,\mu\textrm{m}$ case from figs.\,\ref{fig:3}(e--g).   
It is immediately clear that both the direct fit and the water pole coefficients describe the observed expansion trajectory well.  
The water droplet impact on pole studies differ considerably in the initialization of the sheet formation: droplet impact on a surface versus instantaneous plasma pressure recoil. 
Hence, an imperfect match with the laser-tin data would be expected when using "wp" input and the agreement is in fact surprisingly good. \\ 

\myadded{We next discuss in more detail how the coefficients of the tin radially expanding sheet compare with those of the theoretically derived and validated water sheet expansion in the air upon impact:  }

\begin{itemize} 
\item \myadded{The maximum sheet expansion time for our laser-on-tin experiments is $T_{\textrm{m}}=0.38$ while it is $T_{\textrm{m,p}}\approx 0.43$ for the water droplet impact on a pole \mbox{\citep{wang2023non}}, i.e., $T_{\textrm{m}}/T_{\textrm{m,p}}\approx0.9$. 
This shift to earlier sheet radius apex time could be attributed in part to mass ablation from the droplet surface upon laser impact. 
In fact, previous studies estimated  approximately $10-20\%$ mass loss from ablation \mbox{\citep{HernandezRueda2022,liu_2023_mass}}. 
Such mass ablation would result in a shifted capillary time:  $0.95-0.9\tau_{\textrm{c}}$.  
} 

\item \myadded{We observe a ratio $b_0/b_{\textrm{0,p}}\approx 1.2$ for the coefficient $b_0$ associated with the maximum sheet radius. The differences in both $T_\textrm{m}$ and $b_0$ values are consistent also with a reduced shear stress, i.e.,  energy loss, from a surface-free laser-driven sheet expansion expected to result in a larger maximum sheet radius. Moreover, relative to the  impact on pole, the "over-expansion" of the tin sheet is reminiscent of higher  maximum sheet expansion of \mbox{\citet{lastakowski2014bridging}} for impact on superheated solid surfaces  compared to impact on pole \mbox{\citep[e.g., figure 16 in][]{wang2023non}}.}

\item \myadded{Regarding $b_2$, associated with the deceleration of the rim: the ratio $b_2/b_{\textrm{2,p}}\approx 1.4$, indicates a larger decrease in sheet velocity for the laser-driven expansion compared to that which is pole-impact driven.}

\item \myadded{Finally, the ratio $b_3/b_{\textrm{3,p}}\approx 2.6$ reflects a  larger initial expansion rate for laser-driven tin sheets.} 
\end{itemize}

\myadded{Overall, these distinctions are further reminders of the fundamental differences in  initial impulse conditions causative of the radially expanding sheet formation in the two systems compared here. 
For example, the presence of a bulky central mass after sheet contraction in liquid tin \mbox{\citep{Liu_Bo_phdthesis}} could be consistent with  possible cavitation inside the droplet, hence, compressibility effects on timescales $\sim\tau_{\textrm{pp}}$ \mbox{\citep{reijers2017,liu_2023_mass}}. 
Taken together, these observations further reinforce the need for future theoretical and experimental investigations on  how the initial impulse causative of the radially expanding  sheet formation differ in the laser-on-tin impact versus water droplet impact on a pole of comparable size to that of the impacting drop.\\}

Next, we compare the three thickness profiles in fig. \ref{App:Thickness-Expansion}(a). 
Two of the thickness profiles served as input to obtain the expansion trajectories in fig. \ref{App:Thickness-Expansion}(b), and a third is newly introduced and is derived from the fitted expansion trajectory. 
Given the close match of the "wp" and "tw" expansion coefficients, it is no surprise that the two thickness profiles are quite similar. 
However, these thickness profiles deviated significantly from the profile reported in the work of \citet{liu_2020_mass}. 
We make no claim in the current work as to having found the actual thickness profile. 
Instead we highlight the need for an experimental re-assessment of the thickness profile and note that the underlying differences in the impact conditions, comparing water-pole to laser-impact may be the origin of the discrepancy. 

\subsection{Number of elements}
The time variation of the population of ligaments can be analytically predicted for the water-pole impact system by considering the ratio between the perimeter of the rim bounding the sheet, and the minimal inter-ligament distance, $\lambda_{\textrm{m}}$. 
Here, $\lambda_{\textrm{m}}$ is determined in \citet{wang_bourouiba_2021_growth} from conservation laws and the balance between liquid flowing into the rim and ligaments shedding fragments and considerations of stretching or compression rates of the rim. 
\citet{wang_bourouiba_2021_growth} defines the number of ligaments as (see equation (6.26) in that work) as

\begin{equation}
    N_{\textrm{l}}\frac{2\pi R_{\textrm{s}}(T)}{\Lambda_{\textrm{m}}(T)}=\frac{8\sqrt{3}Q_{\textrm{out}}(T)}{W^{3/2}(T)\beta(T)}=\Theta(T)(\textrm{We}_\textrm{d}/4)^{3/8},\,\,\, \beta(T)=\sqrt{\alpha(T)^2+\frac{5}{2}\alpha(T)-2}.
\label{Eq:C13}
\end{equation}

\noindent Here, $\Lambda_m=\lambda_m/d_0$ is the non-dimensional inter-ligament distance, $Q_{\textrm{out}}$ is the non-dimensional fluid volume shed from the rim per radian per unit of time\,\citep{wang_bourouiba_energypartition}, $W(T)$ is the non-dimensional average ligament width, and $\alpha(T)$ is the ratio between average rim thickness (recall $B(T)\equiv b(T)/d_0$) and average ligament width

\begin{equation}
    W(T)=\alpha(T)B(T).
\label{Eq:C14}
\end{equation}

\noindent In order to calculate $N_{\textrm{l}}$ we need to properly derive $Q_{\textrm{out}}$ and $\alpha(T)$, since the rim thickness has already been expressed for the specific case of the laser-droplet system (see \S\,\ref{IV.3}). We recall the analytical solution of $Q_{\textrm{out}}$ derived in\,\citet{wang_bourouiba_energypartition} (see equation (5.14) therein), which reads

\begin{equation}
    Q_{\textrm{out}}=c\sqrt{\Phi(T)}Y(T)-\frac{\pi}{4}\frac{d}{dT}
    [Y(T)\Psi^2(T)],
\label{Eq:C15}
\end{equation}

\noindent with $c\approx0.51$ a constant, and where $H(R,T)=\Phi(T)/\textrm{We}_\textrm{d}$, $Y(T)=R_{\textrm{s}}(T)/\textrm{We}_\textrm{d}^{1/2}$, and $B(T)=\Psi(T)/\textrm{We}_\textrm{d}^{1/4}$ are the normalized sheet thickness, sheet expansion, and rim thickness functions, respectively, all expressed in $\textrm{We}_\textrm{d}$-independent form. 
Note that $Q_{\textrm{out}}$ explicitly depends on the sheet thickness profile.
We again give a special attention to sheet thickness profile that can be expressed as (see equation (4.3) in \citet{wang_bourouiba_energypartition}): 
\begin{equation}
    \Phi(T)=\frac{\sqrt{6}T}{6 a_3Y(T)^3+\sqrt{6}a_2Y(T)^2T+a_1Y(T)T^2+1/\sqrt{6}a_0T^3}.
\label{Eq:C16}
\end{equation}

\noindent Next, following \citet{wang_bourouiba_2021_growth} (see equation (B10) therein), to good approximation $\alpha(T)$ can be estimated by solving the following equation:

\begin{equation}
    \frac{3}{5}\sqrt{\Psi}\frac{d}{dT}[\alpha\Psi]=\frac{1}{\alpha^{5/2}}\left[2\left( 1-\frac{\pi \dot{Y}\Psi^2}{8Q_{\textrm{out}}}\alpha^2\beta\right)-\alpha^2\beta\right].
\label{Eq:C17}
\end{equation}

\noindent Note that with the rim expansion model, $R_{\textrm{s}}(T)$, and the sheet thickness profile, $\Phi(T)$, we are able to self-consistently calculate both $Q_{\textrm{out}}(T)$ and $\alpha(T)$, from (\ref{Eq:C15}) and (\ref{Eq:C16}), respectively. 
The parameter $\alpha(T)$ is expressed as follows (see equation (C1) in \citet{wang_bourouiba_2021_growth}):

\begin{equation}\label{Eq:C18}
    \alpha(T)=\alpha_0+\alpha_1(T-T_{\textrm{m}})+\alpha_2(T-T_{\textrm{m}})^2,
\end{equation}

\noindent which we substitute in (\ref{Eq:C16}), rearrange terms, and ensure that the prefactor of each power of $T$ equals 0. 
After some algebraic manipulation, the coefficients $\alpha_0$, $\alpha_1$, and $\alpha_2$ are determined. The three sets of coefficients for (\ref{Eq:C17}) from different thickness profiles are: $\left\{\alpha_0=1.10, \alpha_1=0.40, \alpha_2=1.22\right\}_{\textrm{ef}}$, $\left\{\alpha_0=1.10, \alpha_1=0.54, \alpha_2=0.35\right\}_{\textrm{wp}}$, and $\left\{\alpha_0=1.10, \alpha_1=0.36, \alpha_2=1.09\right\}_{\textrm{tw}}$, from empirical fit, water on pole (we employ the published values as inputs), and the profile used in this work, respectively. The derivation of the number of ligaments follows from (\ref{Eq:C13})--(\ref{Eq:C18}). 
These predictions lead to different curves, depicted in fig.\,\ref{fig:5}(f) in the main text.  

\newpage

\end{document}